\documentclass[times,usenatbib]{mn2e}
\usepackage{graphicx,amsmath}
\usepackage{natbib}

\begin{document}

\title[Phase dependent view of Cyclotron lines]{Phase dependent view of Cyclotron lines from model accretion mounds on Neutron Stars} 
\author[Mukherjee and Bhattacharya]
{{Dipanjan Mukherjee $^{1}$, Dipankar Bhattacharya $^{2}$}
\vspace{0.3cm}\\
$^{1}$ IUCAA, Post Bag 4, Pune 411007, India. {\rm Email :} dipanjan@iucaa.ernet.in \\
$^{2}$ IUCAA, Post Bag 4, Pune 411007, India. {\rm Email :} dipankar@iucaa.ernet.in}
\pagerange{\pageref{firstpage}--\pageref{lastpage}} \pubyear{2011}
\maketitle

\begin{abstract}
In this paper we make a phase dependent study of the effect of the distortion of local magnetic field  due to confinement of accreted matter in X-ray pulsars on the cyclotron spectra emitted from the hotspot . We have numerically solved the Grad-Shafranov equation for axisymmetric static MHD equilibria of matter confined at the polar cap of neutron stars.  From our solution we model the cyclotron spectra that will be emitted from the region, using a simple prescription and integrating over the entire mound. Radiative transfer through the accretion column overlying the mound may significantly modify the spectra in comparison to those presented here. However we ignore this in the present paper in order to expose the effects directly attributable to the mound itself. We perform a spin phase dependent analysis of the spectra to study the effect of the viewing geometry.
\end{abstract}


\section{Introduction}\label{intro}
Neutron stars in high mass X-ray binaries have high magnetic fields ($\sim 10^{12}$G) and accrete matter from their companion stars either via stellar winds or by disc accretion. Magnetospheric interaction with the accretion flow causes the matter to be channelled to the magnetic poles, forming accretion columns (see e.g. \citet{ghosh77}, \citet{ghosh78}, \citet{romanova02} and \citet{romanova03}). The infalling plasma, with initial relativistic infall velocities, passes through an accretion shock at a height of a few kilometres from the neutron star surface and then settles down to a gradually slowing subsonic flow \citep{brown98,cumming01}. 

Such X-ray binary systems show characteristic cyclotron resonance scattering features (CRSF) in their spectra resulting from resonant scattering of radiation by electrons in the presence of strong magnetic field (for discussion on theory and observation of cyclotron scattering features see e.g. \citet{harding87}, \citet{araya99}, \citet{araya2000}, \citet{schonherr07} and \citet{mihara07}). In the immediate post shock region the flow velocities are still relativistic ($\sim 0.16c$ for $\gamma=5/3$ gas) and the plasma is optically thin to cyclotron scattering. As the accreted plasma descends and cools, it forms at the base a static mound confined by the magnetic field, and becomes optically thick to cyclotron scattering. Any distortion of the magnetic field in the mound due to pressure from the confined plasma will be reflected in the spectra emitted from the boundary of this region.

The nature and variation of the cyclotron spectra can give important clues regarding the properties of the emission region. Many systems show variations of line energies of the CRSF with the phase of rotation e.g. Vela X-1, Her X-1, 4U 0115+63,GX 301--2 etc. This can be due to the variation of the local magnetic field structure at one or both poles as a line of sight moves across the neutron star. Apart from the spin phase dependence, the cyclotron spectra are also seen to depend on the luminosity state of the system. Some systems like V0332+53 \citep{tsygankov10} show a negative correlation between luminosity and cyclotron line energy while some like Her X-1 \citep{staubert07} show a positive correlation. Such dependence of the line energy with change in accretion rate suggests change of local geometry or magnetic field structure. Some sources (e.g. 4U 1538--52, A 0535+26, V 0332+53 etc) show multiple absorption features with anharmonic separation which can be due to distortion of local field from dipolar magnetic field \citep{nishimura05,nishimura11}. 

In this paper we examine the effect on the cyclotron spectra arising from the distortion of local magnetic field caused by the confined plasma. We consider an accreted mound in static equilibrium confined by the magnetic field at the magnetic pole of a neutron star. We construct the equilibrium solution by solving the Grad-Shafranov equation. We do not consider the effects of continued accretion in this paper. We model the X-ray emitting hotspot as a mound of accreted matter with finite height and no atmosphere. The Grad-Shafranov equation for the accreted matter on the neutron star poles has been previously solved by other authors e.g. \citet{hameury83}, \citet{brown98}, \citet{litwin01}, \citet{melatos01}, \citet{melatos04}, \citet{payne07} and \citet{vigelius08} whose main aim was to study the extent of deformation and stability of the confined mound and also to deduce the effects of magnetic screening on the dipole moment of a neutron star. In this paper we extend this body of work to predict the cyclotron spectra emanating from such mounds.  

We adopt a geometry similar to that used by \citet{hameury83}, \citet{brown98} and \citet{litwin01} and a numerical algorithm similar to the one developed by \citet{mouschovias74} and \citet{melatos04} (PM04) for solving the Grad-Shafranov equation. However our treatment differs from PM04 in several aspects. We work in an axisymmetric cylindrical coordinate system instead of the spherical coordinate system of PM04. We use a polytropic equation of state for the accreted gas instead of the isothermal equation of state of PM04. Finally, we consider the mound to be strictly confined to the polar cap region, while PM04 allowed a significant amount of mass loading outside the polar cap.

We simulate the cyclotron spectra emitted from the accreted mound and perform a phase resolved analysis of the emission. Our main objective in this paper is to perform a phase dependent study of the effects of accretion induced distortion of the local magnetic field on the emergent spectra. In this work we do  not perform a  detailed radiative transfer calculation of CRSF. Instead we use a Gaussian profile for the cyclotron feature originating from each point of the emission region, with the central line energy given by the magnetic field strength at that point, according to the well known relativistic formula given by \citet{sokolov68}. We also incorporate the effects of gravitational bending of light and finite energy resolution of detectors. We generate the resultant spectra by integrating over the entire mound, taking into account the variation of the field strength over the emitting region.

We structure the paper as follows. In Sec. ~(\ref{sec.GSsolve}) we first review the formulation of the Grad-Shafranov equation for the static MHD equilibria. We then outline the numerical algorithm adopted to solve the Grad-Shafranov equation and the test cases for verifying the code. In Sec. ~\ref{sec.results} we discuss the nature of the solutions obtained by solving the Grad Shafranov equation and discuss the range of parameter space within which the valid solution can be obtained. Our results are indicative of the onset of MHD instabilities beyond this boundary.  In Sec. ~\ref{cyclosim} we describe the algorithm used to simulate the spectra from the mound and discuss the results from our simulation of the cyclotron absorption features. In Sec. ~\ref{summary} we summarise the results obtained from the simulations of the spectra and discuss the implications on observations of actual sources. The technical details of the geometrical construct used to compute the spectra are presented in Appendix ~\ref{losgeo}. 

\section{Static MHD equilibria of accreted matter on neutron star poles}\label{sec.GSsolve}
In this work we consider a neutron star binary system where the magnetosphere cuts off the accretion disc at Alfv\'en radius \citep{ghosh77,ghosh78} and matter is accreted on to the polar cap. We will consider a typical slowly spinning neutron star of mass $1.4 M_\odot$, radius $R=10$~km  and magnetic field $\mathbf{B} = 10^{12}$~G (e.g. \citet{dipankar91}). A polar cap of radius $R_p = 1$~km will be considered corresponding to the footprint of dipole field lines that extend beyond a typical Alfv\'en radius of $\sim 1000$~km. Accreted matter is assumed to be confined within the polar cap region. We will consider the accreted matter to form a  degenerate mound of finite height  ($\simeq 100$~m or less)  with a polytropic equation of state. We assume that the mound is in a steady state equilibrium supported by the magnetic field. We work in a cylindrical geometry ($r,\theta,z$) with origin at the base of the polar cap (see Appendix ~\ref{losgeo}) and consider Newtonian gravity with constant acceleration \citep{hameury83,litwin01}.
\begin{equation}
\mathbf{g} = -1.86 \times 10^{14} \left(\frac{M_*}{1.4M_\odot}\right) \left(\frac{R_s}{10 \rm km}\right)^{-2} \mbox{ cm s}^{-2} \mbox{\hspace{1mm}}\hat{\mathbf{z}}
 \end{equation}
The initial magnetic field (when no accreted matter is present) is dipolar. We approximate the dipolar field in the region by an uniform field along z : $\mathbf{B}_{\rm d}=B_0 \boldsymbol{\hat{\textbf{z}}}$. We assume axisymmetry of the polar cap mound and use the ideal MHD equations, which may be cast in the form of the Grad-Shafranov equation. We solve this numerically to find the field and matter density configuration for the static equilibrium solution of the system.


\subsection{Formulation of Grad-Shafranov equation}
For an axisymmetric system, one may decompose the magnetic field into a poloidal and a toroidal part :
\begin{equation}\label{fielddecomp}
\mathbf{B} = \mathbf{B}_p + \mathbf{B}_\theta  = \frac{\boldsymbol{\nabla} \psi \times \hat{\boldsymbol{\theta}}}{r}
\end{equation}
For our work we will assume $\mathbf{B}_\theta = 0$. The function $\psi (r,z)$ is the flux function which at a fixed r and z is proportional to the poloidal flux passing through a circle of radius r (see \citet{kulsrud,biskamp} for more discussion on this). The poloidal components of the magnetic field are
\begin{equation}\label{polB}
\mathbf{B}_r = -\frac{1}{r}\frac{\partial \psi}{\partial z} \mbox{ ; } \mathbf{B}_z = \frac{1}{r}\frac{\partial \psi}{\partial r}
\end{equation}
Using eq. ~(\ref{fielddecomp}) we can write the static Euler equation as 
\begin{equation}\label{newforce}
\boldsymbol{\nabla}p - \rho \mathbf{g} + \frac{\Delta ^2 \psi}{4\pi r^2}\boldsymbol{\nabla}\psi = 0
\end{equation}
where $\Delta ^2$ is the Grad-Shafranov operator : $\Delta ^2 = r\frac{\partial}{\partial r}(\frac{1}{r}\frac{\partial }{\partial r}) + \frac{\partial ^2}{\partial z^2}$. Assuming an adiabatic gas $p={\rm k}_{\rm ad}\rho ^\gamma$ we separate the r and z components \citep{hameury83,litwin01} by the method of characteristics (similar to \citealp{melatos04})
\begin{equation}\label{ch}
p_z - \rho g + \frac{\Delta ^2 \psi}{4\pi r^2} \psi _z=0 \mbox{ ; } p_r + \frac{\Delta ^2 \psi}{4\pi r^2} \psi _r=0 
\end{equation}
where the subscripts indicate partial derivatives. Eliminating $\frac{\Delta ^2 \psi}{4\pi r^2} $ from (\ref{ch}) we get the equation of the integral curve as :
\begin{equation}
dz = -\frac{dr}{\psi _z/ \psi _r} = -\frac{d \rho}{\rho g/c^2_s}
\end{equation}
where $c^2_s=\gamma p /\rho$ is the adiabatic speed of sound. Solving the above two equations, we get $\psi = \mbox{constant}$ (which means the solutions are on constant $\psi$ surfaces) and $gz + \frac{\gamma p}{(\gamma -1)\rho} = f(\psi)$. Here $f(\psi)$ is a $\psi$ dependent constant of integration. Rearranging the terms we can write the density as \citep{hameury83} 
\begin{equation}\label{rhoeq}
\rho = A \lbrack Z_0 (\psi) - z \rbrack ^{\frac{1}{\gamma-1}}
\end{equation}
where $A = \lbrack g (\gamma -1)/(\gamma {\rm k}_{\rm ad}) \rbrack^{\frac{1}{\gamma -1}}$ is a constant.  The function $Z_0(\psi)$ is the mound height profile which defines the vertical height of the mound for an adiabatic gas expressed in flux coordinate ($\psi$) instead of r. The values of $\rho, p$ and their derivatives go to zero smoothly at $z=Z_0(\psi)$. 

Putting eq. ~(\ref{rhoeq}) in eq. ~(\ref{newforce}) we obtain the Grad-Shafranov equation (hereafter G-S) for an adiabatic gas. 
\begin{equation}
\frac{\Delta ^2 \psi}{4\pi r^2} = -\rho g \frac{dZ_0}{d\psi}
\end{equation}
The Grad-Shafranov equation is a coupled non-linear elliptic partial differential equation. We have solved the Grad-Shafranov equation numerically following the algorithm outlined in Appendix ~\ref{secnumGS}.


\section{Results and Analysis of solutions of Grad-Shafranov equation}\label{sec.results}
We have made several runs with different mound height profiles. The solutions show expected behaviour of matter pushing field lines outwards until the tension in the field lines support the gas pressure. The deformation of the field lines increase with larger base pressure and density of the mound  till the solution breaks down after a  threshold density (see Sec. ~\ref{solval}). In this section we discuss the solutions from some sample runs and the range of parameters for which equilibrium solutions can be found.


\subsection{Modelling the magnetically supported accretion mound}\label{sec.modelling_the_mound}
\begin{figure}
	\centering
	\includegraphics[width = 6cm, height = 6cm,keepaspectratio] {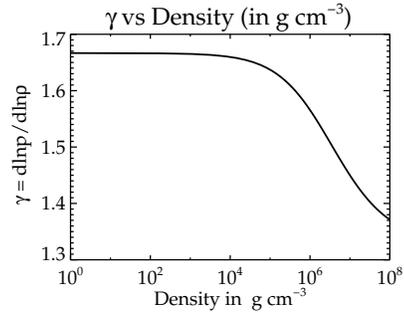}
	\caption{ \small Plot of adiabatic index ($\gamma$) vs density ($\mbox{g cm}^{-3}$). For lower densities $\gamma$ asymptotically converges to 1.6667, which is the value obtained in the non-relativistic approximation. For $\rho \geq 10^8 \mbox{g cm}^{-3}$ $\gamma$ converges to 1.333 which is the value obtained in the ultra-relativistic approximation.}
	\label{gammafig}
\end{figure}

 We will assume a hydrogen poor plasma ($\mu _e =2$) being confined in the mound \citep{brown98}. We restrict the analysis to the gaseous state before ions form a liquid phase. The electrostatic coupling parameter ($\Gamma$) gives a rough estimate whether matter is solid ($\Gamma >> 1$), liquid ($\Gamma \simeq 1$) or gas ($\Gamma < 1$) (e.g. \citet{litwin01})
\begin{eqnarray}
\Gamma &=& \frac{Z^2 e^2}{k_B T}(\frac{4 \pi n}{3})^{1/3} \nonumber \\
&\simeq& 1.1 \left(\frac{Z^2}{A^{1/3}}\right)\left(\frac{\rho}{10^8 \mbox{g cm}^{-3}}\right)^{1/3}\left(\frac{10^8 K}{T}\right)
\end{eqnarray}
Hence mounds of base densities  $< 10^8 \mbox{~g cm}^{-3}$  are considered for this work. To determine the appropriate form of equation of state, we first check if the plasma is non-relativistic ($\gamma = \frac{5}{3}$) or relativistic ($\gamma = \frac{4}{3}$) by evaluating the adiabatic index ($\gamma = \frac{d{\rm ln} p}{d{\rm ln} \rho}$) from the expression of fermionic pressure for degenerate electron gas \citep{chandra_stellar}. As shown in Fig. ~\ref{gammafig}, at $\rho \simeq 10^7 \mbox{g cm}^{-3}$ we have $\gamma \geq 1.4$. For densities lower than this $\gamma$ rises and reaches 1.667 asymptotically. Since, for $\rho \leq 10^7 \mbox{~g cm}^{-3}$, $\gamma$ is significantly higher than 1.33 we model the accreted matter in the mound as a degenerate non-relativistic electron gas with a thermal proton background whose pressure is negligible compared to electron degeneracy pressure. The equation of state for such a system is (values quoted are in cgs)
\begin{equation}\label{eos}
p =  [(3 \pi ^2)^{2/3}\frac{\hbar ^2}{5m_e}]\left(\frac{\rho}{\mu _e m_p} \right) ^{5/3} = 3.122 \times 10^{12}  \rho^{5/3} 
\end{equation}

The plasma is dominated by degeneracy pressure if $\frac{T}{T_F} < 1$,  where $T_F$ is the Fermi temperature : \[T_F = \frac{m_e c^2}{K_B} \lbrack \sqrt{X_F^2+1} -1 \rbrack\] $X_F \!=\! \frac{p_F}{m_e c} \!=\! \frac{1}{m_e c}(\frac{3 h^3}{8 \pi \mu _em_p})^{1/3} \rho ^{1/3}$, $p_F$ being the Fermi momentum.  Observed X-ray continuum from high mass X-ray binaries \citep{coburn04,becker07} indicate that photospheric temperatures in the hotspot are in the range $T \sim 5-10$~keV. The Fermi temperature falls below 10 keV only for a thin layer ($\sim  0.01 Z_0(\psi)$, from eq. ~\ref{rhoeq}) at the top and a major fraction of the mound has Fermi temperature larger than 100keV. Comparison to the temperature profiles obtained including heat transfer effects by \citet{brown98}, shows that the temperature inside the mound remains lower than Fermi temperature at greater depths. Thus modelling the confined accreted matter as a degenerate mound is appropriate.


We have solved the G-S equation for different forms of the mound height profile (e.g. \citet{litwin01,hameury83}) 
\begin{eqnarray}
Z_0(\psi ) &=& Z_{\rm c}(1-(\frac{\psi}{\psi _p})^2) \label{parabolicpro} \\
Z_0(\psi ) &=& Z_{\rm c} \exp(-2\frac{\psi}{\psi _p}) \label{exppro}\\
Z_0(\psi ) &=& Z_{\rm c}(1-(\frac{\psi}{\psi _p})^4) \label{polyn4pro}
\end{eqnarray}
The mound height profile depends on the mass loading function at the accretion disc and redistribution of matter in the shock region for which at present there is no clear knowledge. We resort to evaluating the density by specifying the mound height as a simple function of $\psi$, subject to the constraint : $\rho \rightarrow 0$ as $r \rightarrow R_p$, so that the mound is confined within the polar cap. In most of the analysis we have used eq. ~(\ref{parabolicpro}) which has relatively shallow gradients and helps in speeding up the numerical convergence. 

\subsection{Solutions from the GS-solver}\label{secGSsol}
\begin{figure}
	\centering
	\includegraphics[width = 5cm, height =5cm,keepaspectratio]{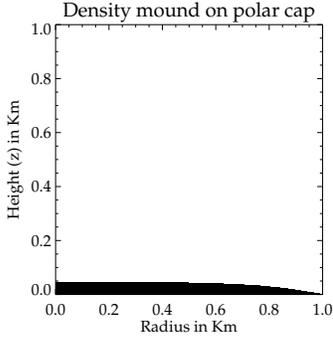}
	\caption{ \small The shape of the accretion mound plotted along r and z axis in equal scale to show the real aspect ratio. The mound is like a thin flat layer on the pole.}
	\label{twoscale}
\end{figure}

\begin{figure}
	\centering
	\includegraphics[width = 7cm, height = 7cm,keepaspectratio] {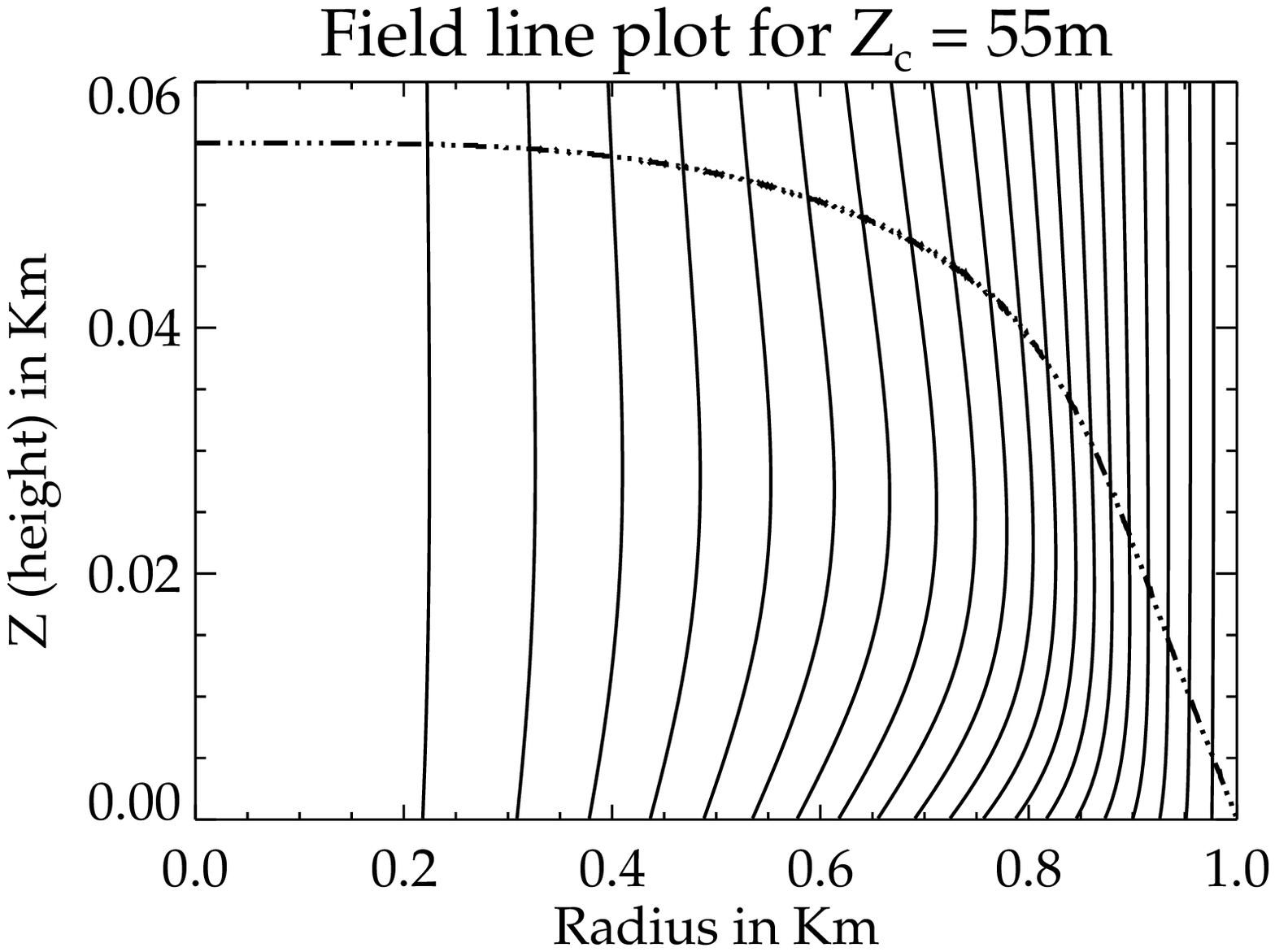}
	\includegraphics[width = 7cm, height = 7cm,keepaspectratio] {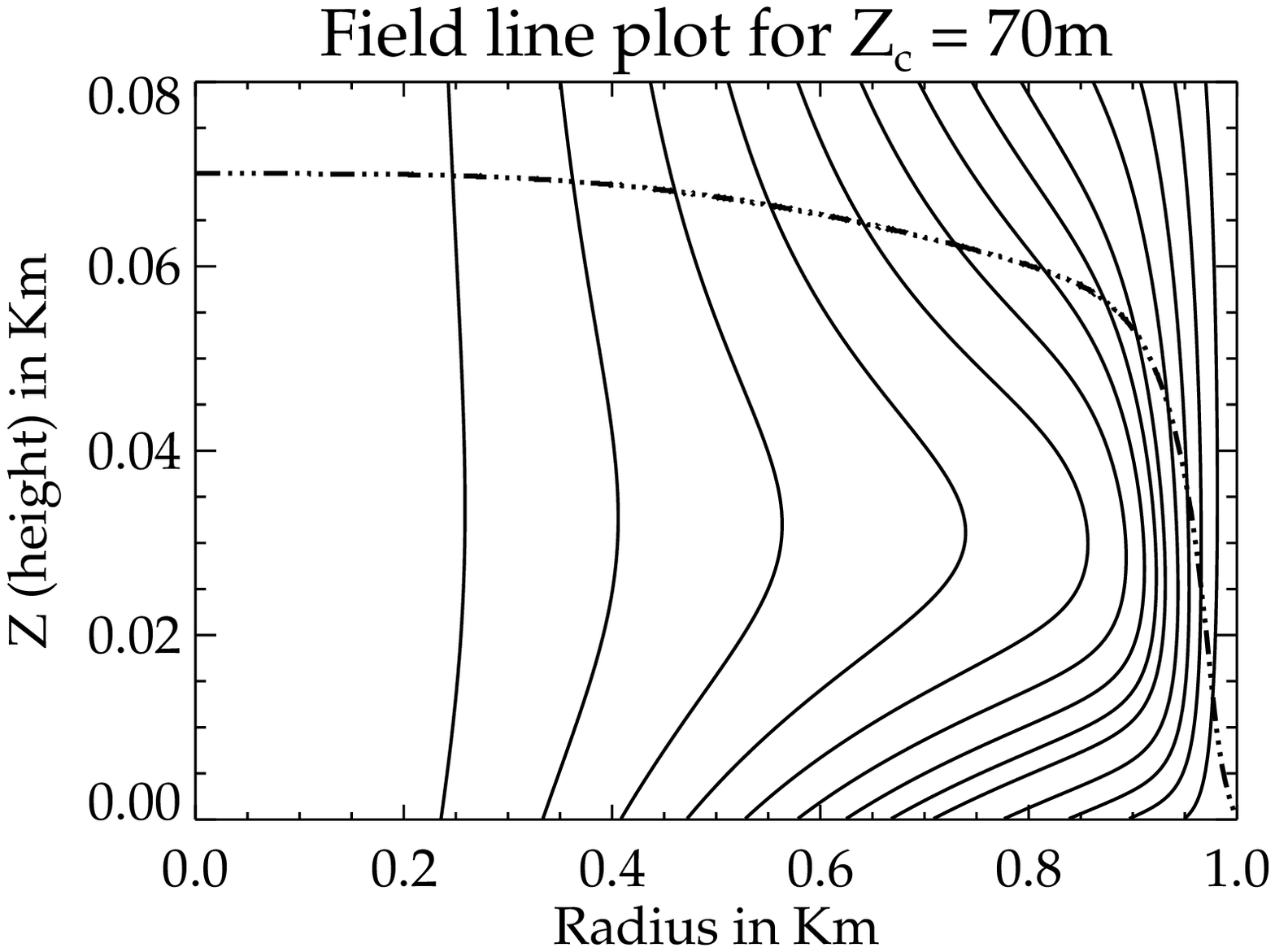}
	\caption{ \small Solution for $Z_c=55$~m (top fig) and $Z_c=70$m (bottom fig), cases (a) and (b) in the text. Solid lines are field lines from G-S solution. The dash-dotted line represents the top of the mound.}
	\label{GSsolfig}
\end{figure}
We discuss here results from two sample runs 

Case (a) : $Z_{\rm c} = 55$~m 

Case (b) : $Z_{\rm c} = 70$~m

with the mound height profile specified by eq. ~(\ref{parabolicpro}) and magnetic field $B=10^{12}$~G. The value $Z_{\rm c}$ was chosen to keep maximum base densities less than $10^8 \mbox{~g cm}^{-3}$ (as discussed in Sec. ~\ref{sec.modelling_the_mound}). 

For case (a) the total mass of the mound is $ \sim 9 \times 10^{-13} M_\odot$ and maximum base density is $\sim 4.7 \times 10^7 \mbox{~g cm}^{-3}$. For case (b) the total mass of the mound is $ \sim 2.13 \times 10^{-12} M_\odot$ and maximum base density is $ \sim 6.8 \times 10^7 \mbox{~g cm}^{-3}$. The mound is in the shape of a flat thin layer on the surface of the star, confined within the polar cap (Fig. ~\ref{twoscale}). Contours of $\psi$ from the solution, which represent the magnetic field lines (as $\mathbf{B} \cdot \boldsymbol{\nabla}\psi = 0$) are plotted in Fig. ~\ref{GSsolfig}. From the figure we see that the field lines are bent to support the pressure of the confined matter. The distortion is more in case (b). Field lines are pushed outwards from the initial configuration resulting in bunching of field lines and increase in field strength.


\subsection{Valid parameter space for existence of solution}\label{solval}
\begin{figure}
	\centering
	\includegraphics[width = 7cm, height = 7cm,keepaspectratio] {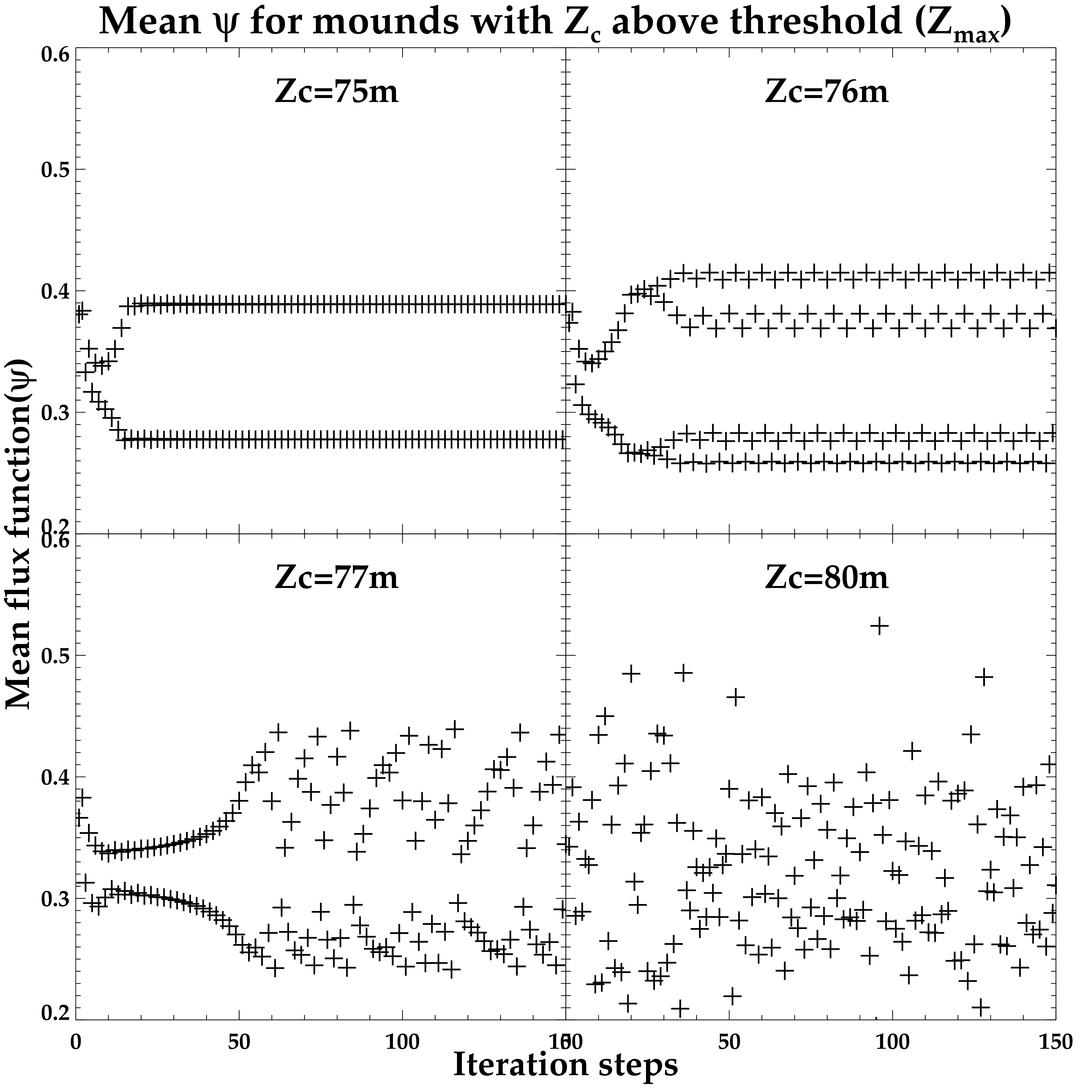}
	\caption{\small Mean $\psi$ as a function of iteration steps for different mound heights above the stability threshold. The mean $\psi$ is seen to oscillate between multiple states. Beyond a certain $Z_{\rm c}$ it passes through different states randomly. }
	\label{meanpsi}
\end{figure}
\begin{figure}
	\centering
	\includegraphics[width = 8.55cm, height = 4cm] {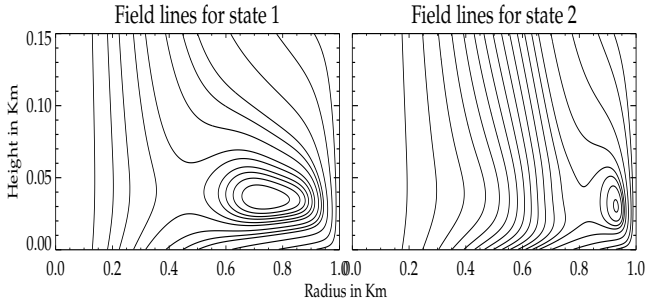}
	\caption{ \small The $\psi$ function at two intermediate iteration steps ($79^{\rm th}$ and $80^{\rm th}$) of the GS-solver for a mound of height 75m. Closed magnetic loops are seen to form which indicate loss of equilibria. At different iteration steps, the $\psi$ passes through states similar to states 1 and 2 depicted here without reaching convergence.}
	\label{closedloops}
\end{figure}
\citet{hameury83} mention in their work that for the configuration of mound height profiles they had considered, no solution was found for field lower than a critical value. We observe the same for different values of magnetic field and different mound height profiles. For a fixed magnetic field we find that for the parabolic profile (eq. ~\ref{parabolicpro}), a stable solution exists for only up to a maximum threshold mound height ($Z_{\rm max}$). For mounds higher than this, the $\psi$ function  keeps oscillating between multiple states with closed magnetic loops during the iteration process and convergence to an unique solution is not reached. For magnetic field $\sim 10^{12}$~G the maximum height of a mound for a stable solution was found to be $Z_{\rm max} \sim 70$m. For a mound higher than this threshold, $Z_{\rm c} = 75$m, the $\psi$ at intermediate steps pass through states similar in nature to states 1 and 2 as shown in Fig. ~\ref{closedloops}. The mean $\psi$ is seen to oscillate between two states as in Fig. ~\ref{meanpsi}. For higher mounds the branches of mean $\psi$ bifurcate to multiple states similar to a pitch-fork diagram. Formation of closed magnetic loops,  also reported in \citet{hameury83}, \citet{melatos04} and \citet{payne07}, appear to indicate loss of equilibria. We have tried  different simulations to check for the existence of solutions beyond the threshold, e.g. higher resolution runs or improving the initial guess solution by starting from a previously converged solution or increasing the radial extent of the grid. However stable solutions were not found for heights greater than the threshold $Z_{\rm max}$.
\begin{figure}
	\centering
	\includegraphics[width = 6.5cm, height = 6.5cm,keepaspectratio] {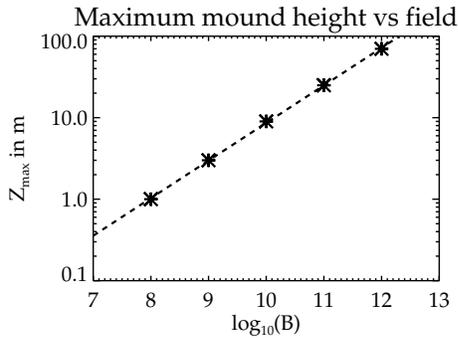}
	\caption{\small Maximum height of the mound ($Z_{\rm max}$) that can be supported by different surface field strengths. Mound height profile of eq. ~(\ref{parabolicpro}) is assumed. The dashed line shows a power-law fit to the points.  Valid solutions can be obtained only for parameters below the line, above which G-S code does not converge.}
	\label{solrange}
\end{figure}

Fig. ~\ref{solrange} shows the  maximum values of $Z_{\rm max}$ (eq. ~\ref{parabolicpro}) for a given field up to which solutions exist. The maximum allowed height ($Z_{\rm max}$) has a power-law dependence on $B$ 
\begin{equation}\label{fieldscaling}
\mbox{log}_{10}(Z_{\rm max}) = -3.676 + 0.461 \mbox{log}_{10}(B)
\end{equation}
where $Z_{\rm max}$ is in metres and $B$ in Gauss. A similar magnetic field to $Z_{\rm max}$ scaling was observed for solution with different mound height profiles e.g for eq. ~(\ref{exppro}) we get $\mbox{log}_{10}(Z_{\rm max}) = -3.79 + 0.46 \mbox{log}_{10}(B)$ and for eq. ~(\ref{polyn4pro}) we get $\mbox{log}_{10}(Z_{\rm max}) = -3.61 + 0.45 \mbox{log}_{10}(B)$  which indicates that this is a generic feature of the Grad-Shafranov equation in the current setup.

Such a scaling relation can be understood approximately by comparing the variation of pressure and magnetic field over different length scales which balance each other. Lateral variation in pressure over scale $R_p$ is balanced by tension from curvature in magnetic field which occurs over a length scale $Z_{\rm max}$. Hence from eq. ~(\ref{rhoeq}), eq. ~(\ref{eos}) and eq. ~(\ref{parabolicpro}) we get
\begin{equation}
\left. \begin{aligned} \rho &\sim A Z_{\rm max}^{3/2} \mbox{ ; } p \propto Z_{\rm max}^{5/2} \qquad \\
\boldsymbol{\nabla}p &\simeq \mathbf{B}\cdot \boldsymbol{\nabla}\mathbf{B} \; \rightarrow \; \frac{p}{R_p} \simeq \frac{B^2}{Z_{\rm max}} \qquad \\
Z_{\rm max} &\propto B^{4/7} \qquad \label{stability}
\end{aligned}
\right \}
\end{equation}
\citet{litwin01} have shown that ballooning instability will disrupt the equilibria if $\Delta \beta > 7.8 R_p/[(\gamma -1)Z_0(\psi)]$, where $\beta$ is the ratio of plasma pressure to magnetic pressure [$p/(B^2/8\pi)$]. Using $ p \sim {\rm k}_{\rm ad} \rho ^{5/3}$ (eq. ~\ref{eos}) and $\rho \sim A Z^{3/2}_{\rm max}$ (eq. ~\ref{rhoeq}) we can write the stability criterion obtained by Litwin as 
\begin{equation}
{\rm log}_{10}(Z_{\rm max}) > -5.1 + \frac{4}{7}{\rm log}_{10}B\label{litwinscaling}
\end{equation}
which is very close to the observed dependence of $Z_{\rm max}$ and $B$ as obtained from our numerical solutions. Thus limit represented by eq. ~(\ref{fieldscaling}) may result from ballooning type pressure driven instabilities where curvature of magnetic field can no longer support the plasma pressure (eq. ~\ref{stability}, eq.~ \ref{litwinscaling}) and the equilibrium solution cannot be obtained. Hence for our analysis of the cyclotron line features in the following section, we restrict ourselves to mounds of height less than 70m for a dipole field of $10^{12}$G. A detailed study of the stability analysis of our solutions will be reported in a future publication (Mukherjee, Bhattacharya and Mignone in preparation).


\section{Cyclotron Resonance Scattering features (CRSF)}\label{cyclosim}
The major part of the mound forms an optically thick medium with cyclotron line formation taking place in a thin layer located at the top surface. The depth of the line forming region may be estimated as $l\sim Z_0 - z$ where $Z_0$ (the mound height profile) is the top height of the mound at a given $r$. From the definition of optical depth and using eq. ~(\ref{rhoeq}) for the density distribution we find the relation between optical depth and thickness of the line forming region as
\begin{equation}\label{taueq}
\tau = \frac{A \sigma}{\mu _e m_p} l^{5/2}
\end{equation}
where $m_p$ is proton mass and $\sigma$ is the scattering cross section. For Thomson scattering ($\sigma = \sigma _{\rm Th}$),  $\tau \simeq 1$ occurs at a depth of $\sim 1.1$~mm and for cyclotron resonance scattering ($\sigma \sim 10^5\sigma _{\rm Th}$)  $\tau \simeq 1$  occurs at 11.3~$\mu$m. Thus cyclotron line formation takes place is a thin layer at the top of the mound. Variation in the local magnetic field at the top of the mound is expected to cause variation in the cyclotron spectra. Modelling of the cyclotron resonance scattering features (hereafter CRSF) taking into account the contribution of different parts of the mound to the line of sight is presented in the following section.


\subsection{Modelling cyclotron spectra}\label{modelling_cyclotron_spectra}
The  emission profile from a point on the mound depends on the strength and direction of the magnetic field and the angle between the emergent radiation and the local normal to the mound surface, which vary with position due to curvature of the field lines and the shape of mound surface respectively. We divide the mound into small area elements ($\Delta A_{r_i,\theta _j}$ in a ($r,\theta$) grid in cylindrical coordinates on the polar cap) and find the local magnetic field vector and local normal to the mound for each element, assuming them to constant over the area element. The resultant cyclotron spectrum is constructed by integrating the weighted contribution of emission from all parts of the mound towards a line of sight (hereafter los)
\begin{equation}
F= \displaystyle\sum _{\rm i,j}I(\theta_{\alpha l})\Delta A_{r_i,\theta _j}\cos \theta _{\alpha l}\label{fluxfunction}
\end{equation}
 $I$ is the normalised intensity at the mound surface and the angle $\theta _{\alpha l}$ (eq. ~\ref{thetal}) is the angle between the direction of emission at the mound surface and the local normal to the surface. For integrating the intensity over the mound we have chosen a radial grid with resolution equal to or higher than the resolution in the radial direction of the G-S solution.

The cyclotron line energy is given by 
\begin{equation}\label{enform}
E_n = nE_{c0}\sqrt{1-u}\left(1-\frac{n}{2}\left(\frac{E_{c0}}{\rm 511 keV }\right)\sin ^2 \theta _{\alpha b}\right)
\end{equation}
where $n=1,2,3...$ is the order of the harmonic, $E_{c0}=11.6B_{12}$ in keV, $\theta _{\alpha b}$ is the angle between the direction of emission and local magnetic field (eq. ~\ref{thetab}) and $u=r_s/r$, $r_s$ being the Schwarzschild radius. The factor $\sqrt{1-u}$ gives the gravitational redshift of the line.

 Eq.~(\ref{enform}) is correct to second order in the small parameter $E_{c0}/({\rm 511 keV})$, and is adequate for the field strengths we consider in this work. In our studies we consider only the effect of the fundamental $n=1$ line. The accurate dependence of the width and depth of the CRSF on the angle $\theta _{\alpha b}$ can be obtained by solving radiative transfer equations in the mound. This is beyond the scope of the present work and will be addressed in a future publication (Kumar et al in preparation). For our present work we model the cyclotron feature by a Gaussian profile with line centres from eq. ~(\ref{enform}) and model the dependence on the angle $\theta _{\alpha b}$ of the line width $\Delta E/E_c$ and the relative depth by interpolating from the results presented in \citet{schonherr07} for the slab 1-0 geometry.

We incorporate effects of gravitational light bending (eq. ~\ref{lightbend}) following the approximate formula given by \citet{beloborodov02}. For the intrinsic intensity profile we use a form $I(\theta _{\alpha l})=I_0+I_1\cos \theta _{\alpha l}$, but set $I_1=0$ for most of our analysis. To simulate the finite energy resolution of the detectors, we convolve the spectra with a normalised Gaussian with the standard deviation a fraction $f$ of the local energy  
\begin{equation}
W(E,E') = \frac{1}{\sqrt{2\pi} \sigma}\exp\left(\frac{-(E-E')^2}{2 \sigma ^2}\right) ; \, \sigma \sim f E' \label{sigmaf}
\end{equation}
Detectors currently used for observations in X-ray astronomy usually have a energy resolution of 10\%-20\% ($f\simeq 0.1-0.2$). We carry out the above analysis at different phases of rotation of the neutron star to perform a phase dependent study of the spectra. The nature of the spectra also depends on the relative orientation of the mound (inclination angle $\eta _p$)  and the los (angle $i$) with respect to the spin axis of the neutron star, which are treated as free parameters (see Appendix ~\ref{losgeo} for details on the geometry).

\subsection{Results : cyclotron spectra from a single mound}\label{single_mound}
We have carried out simulation runs for the cyclotron spectra using the solutions of the magnetic field obtained from the GS-solver. For most of our analysis in this section we use the solution with profile eq. ~(\ref{parabolicpro}). Although the shape and nature of the CRSF will change for different profile functions we can draw some general conclusions about the dependence of the cyclotron spectra on the local magnetic field. 
\begin{figure}
	\centering
	\includegraphics[width = 8cm, height = 8cm,keepaspectratio] {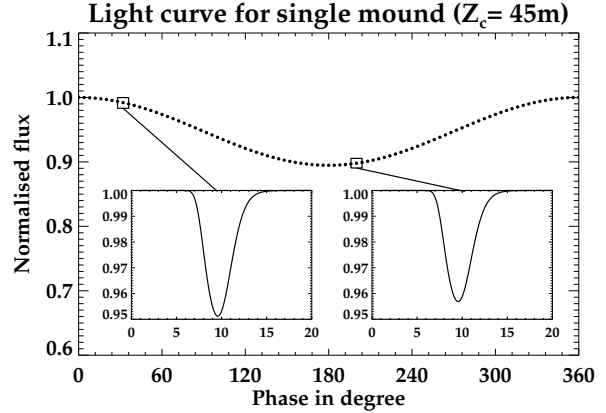}
	\caption{ \small The light curve of one pole mound for $Z_{\rm c} = 45$~m, $\eta _p=10^\circ$ and $i=30^\circ$ (Appendix ~\ref{losgeo}). Insets show the spectra at the two rotation phases marked. No significant variation of line energy is seen. A detector resolution of 10\% was assumed (eq. ~\ref{sigmaf}).}
	\label{spectra_45m}
\end{figure}

\begin{figure}
	\centering
	\includegraphics[width = 7cm, height = 9cm,keepaspectratio] {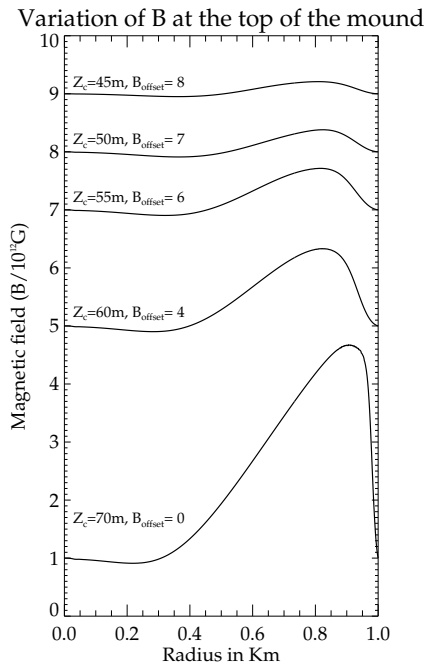}
	\caption{ \small The variation of the strength of the magnetic field with radial distance at the top of the mound, for different $Z_{\rm c}$. The plots are offset by an amount ($B_{\rm offset}$) for clarity. The maximum magnetic field at the top is several times the surface dipole field due to distortion from pressure of accreted matter.}
	\label{btopvar}
\end{figure}

We first consider the case of emission from a single hotspot at one of the poles. Fig. ~\ref{spectra_45m} shows the light curve from a mound of height $Z_{\rm c} = 45$~m at a pole with inclination $\eta _p=10^\circ$ and los $i=30^\circ$ (See Appendix ~\ref{losgeo} for definitions of $i$ and $\eta _p$). The inset plots show the cyclotron spectra convolved with a Gaussian with $f=0.1$ (eq. ~\ref{sigmaf}), at two rotation phases $ 32^\circ \mbox{ and  } 200^\circ$. Although the magnitude of the field at the top of the mound varies by $\sim 27\%$ in this case (Fig. ~\ref{btopvar}), the line centre of the CRSF shows less than 0.2 \% change about a mean $\sim 9.6$~keV. As the continuum emission is assumed to be isotropic and uniform, and also since gravitational bending redirects the light rays in directions well away from straight trajectories, all parts of the mound contribute towards a given line of sight at all phases. This gives a resultant averaged spectrum with very little phase dependence of the CRSF from a single pole.
\begin{figure}
	\centering
	\includegraphics[width = 8cm, height = 9cm,keepaspectratio] {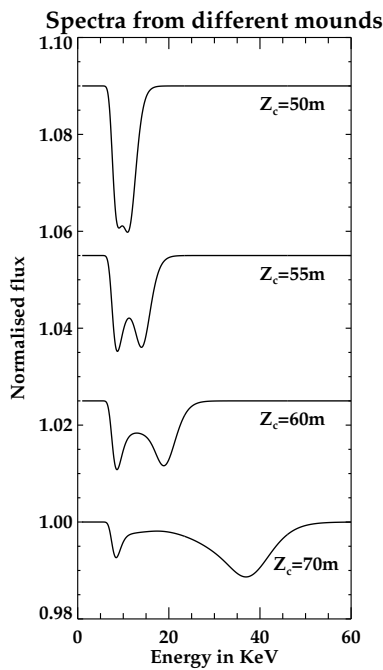}
	\caption{ \small  CRSF from mounds of different heights. The spectrum for each mound shows two CRSF fundamentals at two distinct energies, corresponding to the undistorted dipolar field and the region with large distortion of the magnetic field respectively (see Fig. ~\ref{btopvar}).}
	\label{multispec}
\end{figure}

For mounds with large distortion of surface magnetic field (e.g. for $Z_{\rm c} = 70$~m, the field at the edge rises to $\sim 4.6$ times the original dipole value (see Fig. ~\ref{btopvar}). We find two distinct CRSF fundamentals at two distinct energies (Fig. \ref{multispec}). The feature at a lower energy, for a field $\sim 10^{12}$~G, originates near the centre of the mound while that at a higher energy arises in the high field regions near the periphery of the mound (Fig. ~\ref{btopvar}). We emphasize that this multiple-featured spectrum is a result of the variation of the local field strength and does not represent multiple harmonics, as only $n=1$ features have been included in our computation. The energy ratio of these features may therefore be arbitrary and not follow a harmonic relation.
\begin{figure}
	\centering
	\includegraphics[width = 8cm, height = 8cm,keepaspectratio] {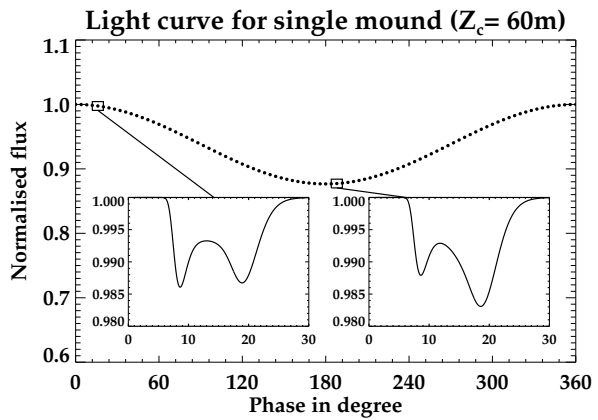}
	\caption{ \small  The light curve of one pole mound with $Z_{\rm c} \sim 60$~m, $\eta _p=10^\circ$ and $i=30^\circ$ (Appendix ~\ref{losgeo}). Insets show the spectra at two the rotation phases marked. Multiple CRSF are seen with variation of the relative depth with phase, but no phase dependence of the line energies. A detector resolution of 10\% was assumed (eq. ~\ref{sigmaf}). }
	\label{spectra_60m}
\end{figure}

The line centres of the individual peaks show little variation with phase but the relative depth of the peaks depend on the viewing geometry and the phase angle. Fig. ~\ref{spectra_60m} shows the light curve and spectra (with $f=0.1$ in eq. ~\ref{sigmaf}) from a mound $Z_{\rm c} \sim 60$~m with $\eta _p=10^\circ$ and $i=35^\circ$, where the relative depth of the CRSF vary with phase.  The degree of their variation depends on the relative orientation of the pole and the los, occurring for a certain range of viewing angles ($25^\circ \leq i \leq 40^\circ$ for a pole at $\eta _p=10^\circ$ in the present case). Two distinct CRSF are observed for all mounds of appreciable magnetic field distortion at the top and all viewing geometries, which shows that it is a generic feature of the cyclotron lines originating from the mound. 
\begin{figure}
	\centering
	\includegraphics[width = 6cm, height = 6cm,keepaspectratio] {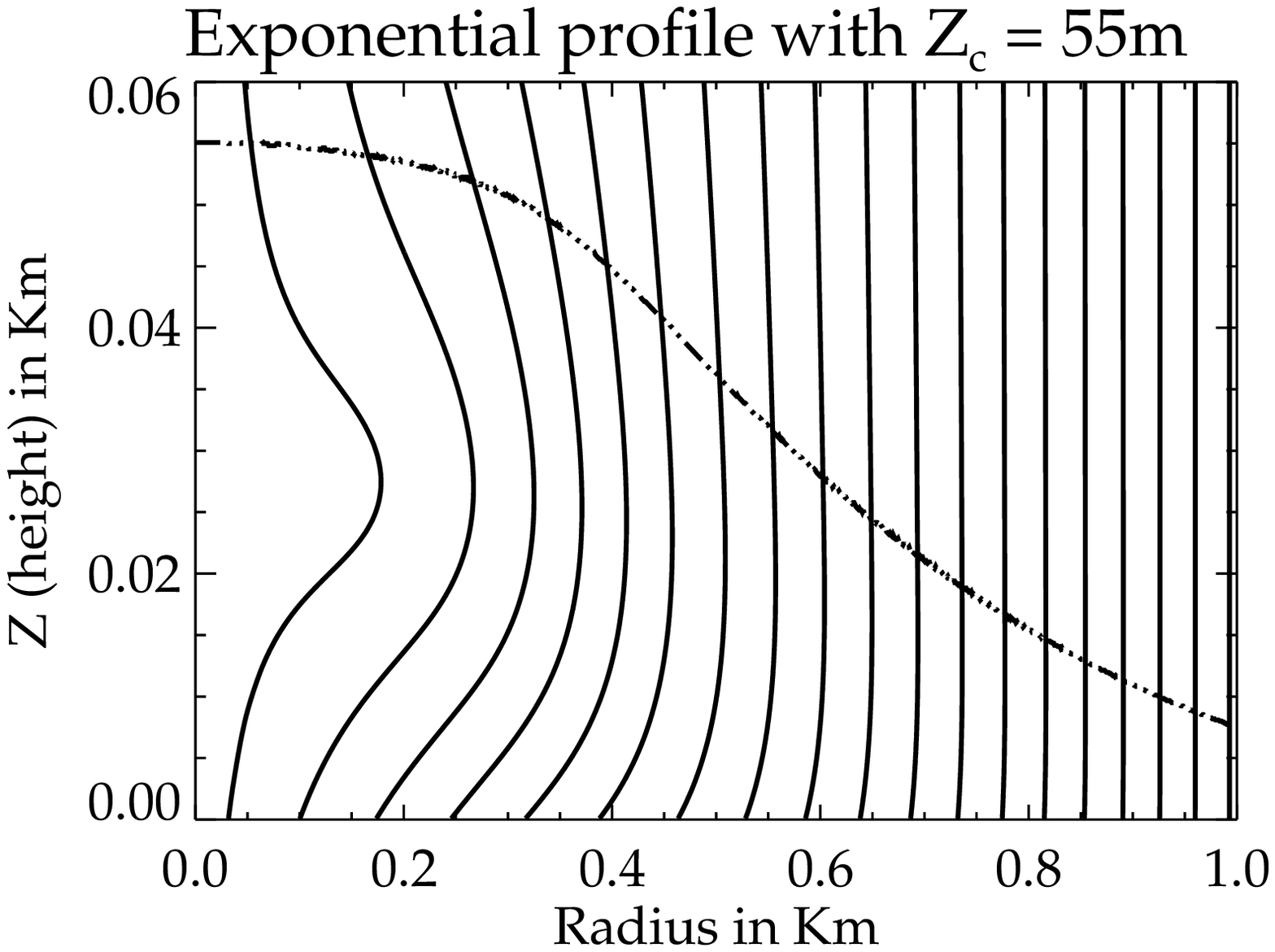}
	\includegraphics[width = 6cm, height = 6cm,keepaspectratio] {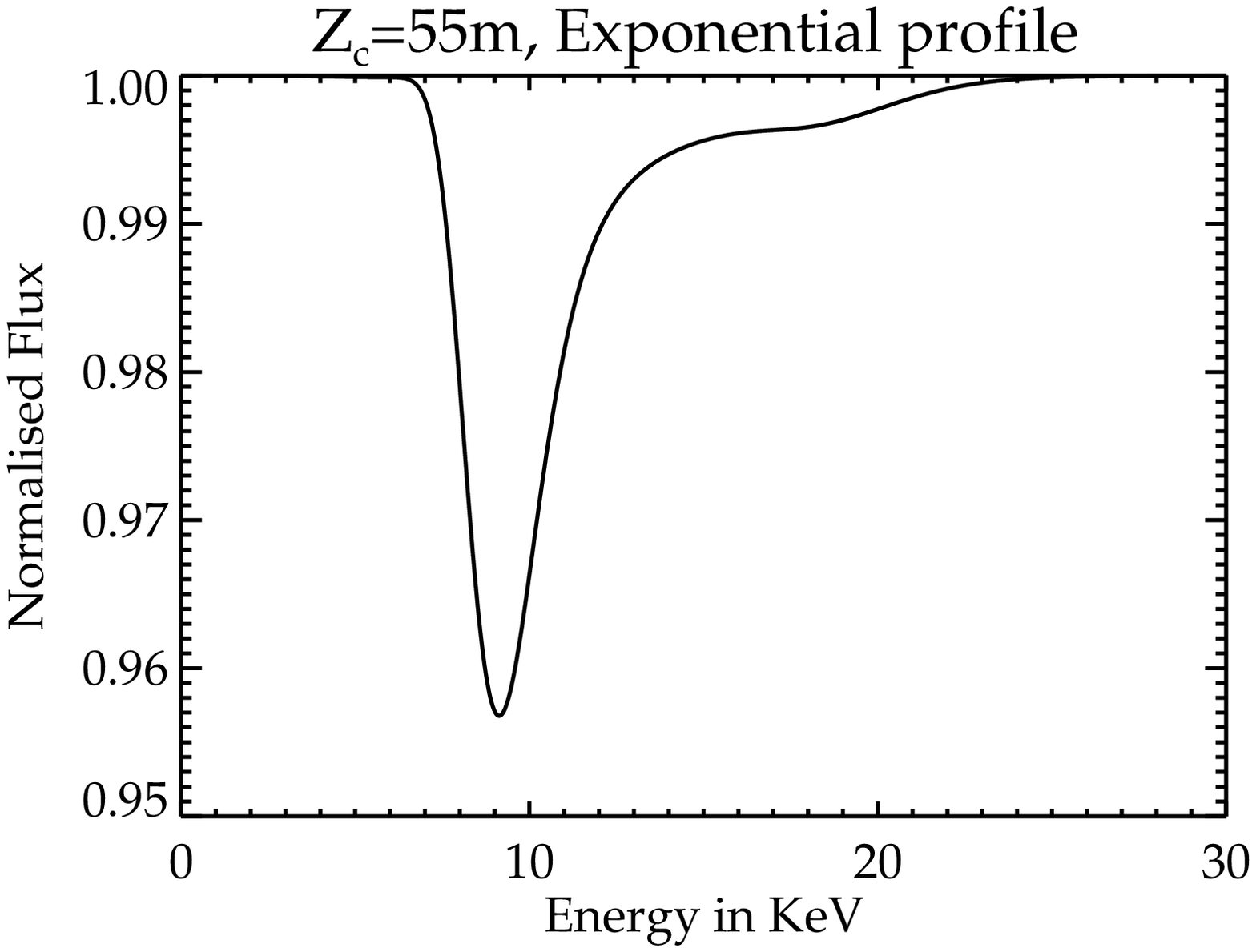}
	\caption{ \small Magnetic field line structure (upper panel) for a mound with $Z_{\rm c} \sim 55$m with an exponential mound height profile eq. ~(\ref{exppro}) and the spectrum emitted (lower panel). The spectrum is dominated by area elements near the polar cap edge (corresponding to the region of undistorted field in this case) as they span larger areas. The shallow feature at $\sim 18$ keV corresponds to the region of large field distortion closer to the axis.}
	\label{spectra_55mexp}
\end{figure}

The cyclotron features are dominated by the field structures near the periphery of the mound as these regions have a larger emitting area.  For G-S solutions with mound height profile as in eq. ~(\ref{parabolicpro}), the regions of higher magnetic field distortion occur near the edge of the polar cap. Hence we observe two CRSF of comparable depths for all mounds with profile eq. ~(\ref{parabolicpro}). However, if we use an exponential profile (eq. ~\ref{exppro}) which falls off sharply with $r$, regions of larger field distortion occur much closer to the axis (Fig. ~\ref{spectra_55mexp}), and the CRSF is dominated by the undistorted field in the outer region. A shallow feature at $\sim 18$~keV is contributed by the higher field regions nearer to the axis.  This shows that the CRSF is heavily influenced by the geometry and distribution of the local field as well as the structure of the mound. 


\subsection{Effects of finite energy resolution of detectors}\label{detector_resolution}
\begin{figure}
	\centering
	\includegraphics[width = 8cm, height = 9cm,keepaspectratio] {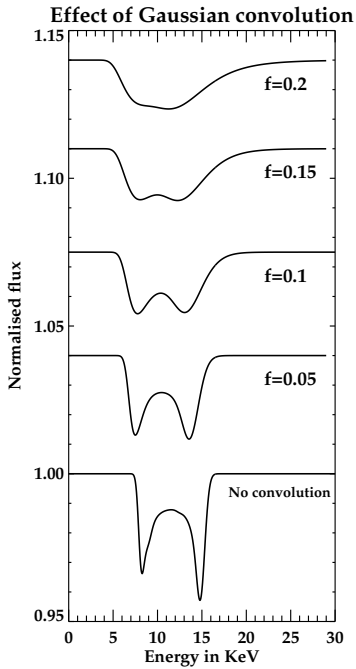}
	\caption{ \small The effect of finite detector resolution modelled by convolving the spectra with a Gaussian function. Results for different values of $f$ (eq. ~\ref{sigmaf}) are shown. The double CRSF nature of the spectra disappear for $f \geq 0.2$ corresponding to a 20\% energy resolution of detector.}
	\label{energyres}
\end{figure}
The finite energy resolution of the detector can make the two absorption features indistinguishable if closely placed. In Fig. ~\ref{energyres} we show the effect of Gaussian convolution with different values of $f$ (eq. ~\ref{sigmaf}). For $f=0.2$, the two peaks are indistinguishable. For mounds of lower height (e.g. $Z_{\rm c} = 45$~m) the two CRSF become indistinguishable at a much smaller $f$. Thus the finite energy resolution of the detector can often mask the effects of the internal magnetic field structure on the CRSF. However for mounds of higher height, the two CRSF have centres far enough to be distinguishable even with $f=0.2$. 

\subsection{Emission from two anti-podal poles}\label{anti-podal_poles}
\begin{figure}
	\centering
	\includegraphics[width = 8cm, height = 8cm,keepaspectratio] {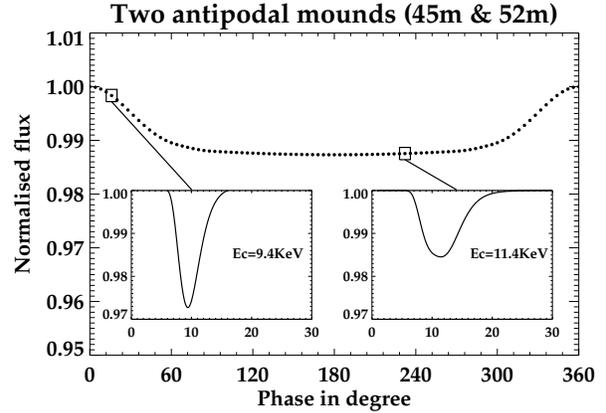}
	\caption{ \small The light curve for two anti-podal mounds, $Z_{\rm c} \sim 45$~m and 52~m, with $\eta _p=30^\circ$ and $150^\circ$ respectively, and $i=30^\circ$ (Appendix ~\ref{losgeo}). Insets show the spectra at the two rotation phases marked. The CRSF shows 20\% change in line energy with phase. A detector resolution of 15\% was assumed (eq. ~\ref{sigmaf}). }
	\label{twopole}
\end{figure}
Many systems show variation of the cyclotron energy with the phase of rotation of the neutron star e.g. $\Delta {\rm E}_{\rm cyc}/{\rm E}_{\rm cyc} \sim 10\%$ for Vela X-1 \citep{kreykenbohm02}, $ \sim 20\%$ for 4U0115+63 \citep{heindl04,baushev09}, $ \sim 25\%$ for Her X-1 \citep{klochkov08_herx1}, GX 301-2 \citep{kreykenbohm04}, and  $ \sim 30\%$ for Cen X-3 \citep{suchy08,burderi2000}. From our simulations of CRSF for emission from a single hotspot we find very little variation of the line energy with rotation phase. However if emission is considered from two antipodal hostpots at opposite poles with slight difference in mound height due to asymmetric accretion, then the line centre of the resultant CRSF show a stronger phase dependence. Fig. ~\ref{twopole} shows the light curve for a neutron star with two anti-podal poles at $\eta _p\sim 30^\circ \mbox{ and } 150^\circ$, having mounds of height $Z_{\rm c} \sim$ 45~m and 52~m respectively, and an observer at inclination $i=80^\circ$. The simulation is carried out by assuming uniform and isotropic continuum intensity normalised to 1 at both poles, which may not be valid in reality due to difference in accretion rates at the two poles. However for small differences in the mound heights considered here, we ignore the differential luminosity effect and  draw some general conclusions about the behaviour of the cyclotron line energies. The spectra are convolved with a Gaussian function with $f=0.15$ (eq. ~\ref{sigmaf}). 

The light curve is not sinusoidal unlike the case of a single pole (Fig. ~\ref{spectra_45m} and Fig. ~\ref{spectra_60m}). The line energy of the CRSF varies by $\Delta E_c \sim 20 \%$ over a full rotation cycle. The spectrum from the pole nearer to the los dominates the CRSF. The observed variation in the line energy of course depends on the viewing geometry defined by the angles $i$ and $\eta _p$. A proper evaluation of the spectra for a two pole case would require the knowledge of the accretion rate and local temperature of the hotspots, but it may be concluded that in the presence of multiple hotspots, the CRSF line energy will show a significant phase dependence.

\section{Discussion and conclusions}\label{summary}

We have modelled the structure of the accretion mound by solving for the static magnetic equilibria described by the Grad-Shafranov equation. For mounds of total accreted mass $\sim 10^{-12} M_\odot$ (Sec. ~\ref{secGSsol}) there is appreciable distortion of the magnetic field. We have found that for a given surface field strength, stable solutions to the G-S equation are not found for mounds of height higher than a threshold (Sec. ~\ref{solval}). Beyond the threshold, field lines with closed loop configurations are formed, and may  indicate the onset of pressure driven MHD instabilities. So we restrict our analysis to mound heights for which equilibrium solution is obtained.

Using the mound structure obtained from the Grad-Shafranov equation, we have simulated the cyclotron spectra which will originate from its mound surface. We have integrated the emission from different parts of the mound to find the resultant spectra and have performed a phase dependent study of the spectra. We have assumed a Gaussian model for the CRSF fundamental and have incorporated effects of gravitational bending of light and finite energy resolution of the X-ray detectors (Sec. ~\ref{modelling_cyclotron_spectra}). From our analysis of the phase dependent spectra (Sec. ~\ref{single_mound}, Sec. ~\ref{detector_resolution} and Sec. ~\ref{anti-podal_poles}) we can draw some general conclusions :
\begin{enumerate}
\item
\begin{figure}
	\centering
	\includegraphics[width = 6cm, height = 6cm,keepaspectratio] {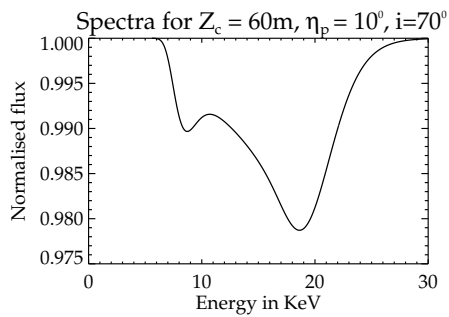}
	\caption{ \small The CRSF from a mound of height $Z_{\rm c} = 60$~m with mound height profile eq. ~\ref{parabolicpro}, $\eta _p=10^\circ$ and $i=70^\circ$ (see Appendix ~\ref{losgeo}). The undistorted field produces the small feature at $\sim 8$~keV. The distorted field near the polar cap edge is the dominant contributor to the CRSF, producing a dip at $\sim 18$~keV.  This  shows that field inferred from CRSF line energy may not depict the surface dipole field, but the field distorted due to accreted matter.}
	\label{spectra_60m_I70}
\end{figure}
{\bf CRSF need not represent the dipole field :} The CRSF are heavily influenced by the distortion of the local field caused by the pressure of the confined accreted matter. The line energy of the CRSF may not always represent the intrinsic dipole magnetic field.  As shown in Fig. ~\ref{spectra_60m_I70}, the small dip at $\sim 8$~keV is the redshifted cyclotron feature due to the surface field of $10^{12}$~G, which may not be observed in practice in the presence of noise and poor detector resolution. The CRSF at $\sim 18$~keV resulting from the distorted field will be the dominant feature in the spectrum. For mounds of height $Z_{\rm c} \sim 70$~m the maximum distortion in the field can be as large as 4.6 times the dipole value.  

\item
{\bf Multiple, anharmonic CRSF :} From our simulations we observe Multiple CSRF fundamentals from a single mound with considerable field distortion (see Fig. ~\ref{multispec}, Fig. ~\ref{spectra_60m}). A real spectrum  will also contain multiple CRSF harmonics which will add to the complexity of the spectrum.  Several HMXBs show multiple CRSF where the line centres have anharmonic separation e.g. $E_c \sim 22$ keV, 47 keV for 4U 1538-52 \citep{rodes09}, $\sim$ 23 keV, 51 keV for Vela X-1 \citep{kreykenbohm02}, $\sim$ 45 keV, 100 keV for A 0535+26\citep{kendziorra94,caballero07}, $\sim$ 26 keV, 49 keV, 74keV for V 0332+53 \citep{makishima90,katja05}, $\sim$ 36 keV, 63 keV for EXO 2030+375 \citep{reig99,klochkov08_EXO}.  More on the properties of these sources can be found in the reviews by \citet{mihara07} and \citet{lutovinov08}. 

Anharmonic line spacing due to intrinsic non-dipolar field has been discussed by \citet{nishimura05}. Our results show that strong non-dipolar structure can be generated in the accretion process itself. 

\item
{\bf Effect of detector resolution :} The detectability of multiple cyclotron features depend on the energy resolution of the detector. For detectors with poor energy resolution, the multiple fundamental features will be masked as shown in Fig. ~(\ref{energyres}). However for mounds with large field distortion ($Z_{\rm c} \sim 60-70$m) multiple absorption features will still be observed with detectors that have energy resolution of $\leq 20\%$.  

\item
{\bf CRSF phase dependence - one pole : } We find no appreciable spin phase dependence of the line energy of a given CRSF from a single mound (Fig. ~\ref{spectra_45m}) despite there being considerable local variation of magnetic field. This may be attributed to the fact that all parts of the mound contribute towards a given line of sight at any phase due to effects of gravitational light bending and assumption of isotropic local emission. However we find that the relative depth of the multiple CRSF depends on phase and viewing geometry as shown in Fig. ~\ref{spectra_60m}.

\item
{\bf CRSF phase dependence - two pole :} Line energy variation is however observed if emission from both poles with slightly different mound heights are considered (Sec. ~\ref{anti-podal_poles}).  For emission from two anti-podal mounds of height $Z_{\rm c} \sim 45$~m and $\sim 52$~m the CRSF line energy varies by 20\% during one spin cycle, similar to what is observed for many sources.  Thus we conclude that emission from multiple accretion mounds will result in strong variation of the line energy of the CRSF with spin phase.

\item
{\bf Dependence on mound structure :} CRSF are dominated by field structure at the mound periphery due to the larger physical area of this region. Different structure and density distribution of the mound would cause different distributions of the local field. Fig. ~\ref{multispec} and Fig. ~\ref{spectra_55mexp} show the difference in the CRSF from a mound of the same maximum height ($Z_{\rm c} = 55$~m) but different mound height profiles. The strong dependence of the spectra on the structure and size of the mound suggests that with variation in the accretion rate, the observed line profiles and energies may change, contributing to a luminosity dependence of the spectrum. 

 Some sources e.g. Her X-1 \citep{staubert07} and A 0535+26 \citep{klochkov11},  show a positive correlation between luminosity and line energy.  In our picture, this can be attributed to the presence of a strong non-dipolar component in the field due to an increase in mass of the mound resulting from an increase in accretion rate. Our simulations show that for a small change of mound size we have an appreciable change in maximum magnetic field at the top of the mound (Fig. ~\ref{btopvar}), e.g. between $Z_{\rm c}$ of 55~m and 60~m, the maximum field at the top of the mound changes by $\sim 36$\%. Since we work in the static limit and do not consider accretion rate as a parameter in our simulations, we cannot directly probe the luminosity dependence of the spectrum. However we can conclude that small changes in height of the mound can result in significant changes in the magnetic field inferred from the CRSF.

\item
{\bf Effect of anisotropic continuum : } We have also carried out runs with mildly anisotropic continuum intensity profiles : $I (\theta _{\alpha l}) = I_0(1+\cos\theta _{\alpha l})$ and have found no appreciable phase dependence of the line energy of a one pole CRSF although the percentage modulation of the flux was larger. However if continuum is highly anisotropic, e.g. $I = \left(\sin 2 \theta _{\alpha l}/(2\theta _{\alpha l})\right) ^2$ then the resultant spectra are found to be phase dependent. However such strong beaming may not be realistic in the context of HMXB pulsars.

\item
{\bf Screening effect from overlying column :} Effects of partial screening of the mound by an atmosphere of accreted matter have not been considered in our simulations. An optically thick blanket of settling plasma can screen a fraction of the mound depending on the viewing geometry. The effect of screening can be estimated approximately by using the velocity profile of the settling flow from  \citet{becker07} : $v(z) = 0.49c \sqrt{z/R_p}$, for a neutron star of mass $\sim 1.4M_\odot$ and radius $\sim 10$km, mean plasma temperature $\sim 10$keV and magnetic field $\sim 10^{12}$G. For an accretion rate $\dot{M}\sim 10^{16} {\rm g s}^{-1}$ and a mound of height $Z_{\rm c} \sim 70$m (maximum allowed $Z_{\rm c}$ for eq. ~\ref{parabolicpro}) we get the Thomson optical depth as $\tau \sim 1.57 \times 10^{-5} \ell$, where $\ell$ is the line element along the los through the accretion column. For a hotspot of radius $R_p \sim 10^5$cm, we see that regions near the axis will be optically thick along a radial line from the axis (in local cylindrical geometry of the mound). The degree of screening of different parts of the mound will vary with spin phase, resulting in phase dependent spectra.
\end{enumerate}

Thus we conclude from this work that the distortions in local magnetic field due to confinement of accreted matter at the polar cap has considerable influence on the phase resolved spectra from accreting binary neutron star systems. Current observations often have poor count statistics and involves averaging in phase due to which many of the details in the spectra may be lost. Future missions like ASTROSAT (Agarwal et. al. 2005 \nocite{agarwal05}, \citet{rao09}) with higher sensitivity and better energy resolution can help us analyse the spectra in greater detail and provide us clues to the nature, shape and geometry of the accretion mound. A more detailed analysis including dynamic simulations of the emitting region and full radiative transfer would help us investigate effects like line energy-luminosity correlation, anharmonic separation of the CRSF energies, asymmetric shape of the CRSF etc.


\section{Acknowledgement}
We thank CSIR India for Junior Research fellow grant, award no 09/545(0034)/2009-EMR-I. We are very thankful to Andrea Mignone for his help and suggestions in setting up simulation runs with PLUTO. We  thank Dr. Kandaswamy Subramanian, Dr. Ranjeev Misra and Sandeep Kumar from IUCAA, for useful discussions and suggestions during the work, and IUCAA HPC team for their help in using the IUCAA HPC where most of the numerical computations were carried out. We also would like to thank Dr R\"{u}diger Staubert for his suggestions in improving the text, and also the anonymous referee for his valuable suggestions and comments.


\appendix

\section{Solving the Grad-Shafranov equation}\label{secnumGS}
\subsection{The numerical algorithm}

We express the G-S equation in non-dimensional form by scaling the physical parameters with $L_0=R_p=1\mbox{km}$, $\rho _0 = 10^6 \mbox{~g cm}^{-3}$, $B_0 = 10^{12}$~G. The flux function is normalised to the value at $r=R_p$ : $u=\psi/\psi _p$ ($\psi _p = \frac{1}{2}B_0 R^2_p$). We perform a variable transformation $x=r^2$ and solve the G-S equation on a $(x-y)$ grid which are the normalised ($r^2-z$) coordinates. The result is then transformed back to the ($r-z$) grid by spline interpolation. With this transformation and scaling, the G-S equation for an adiabatic gas reads as
\begin{equation}
4x\frac{\partial ^2 u}{\partial x^2} + \frac{\partial ^2 u}{\partial y^2} = -{\cal C}x(Y_0 - y)^\alpha \frac{dY_0(u)}{du}
\end{equation}
where ${\cal C}=16\pi Ag L^{1+\alpha}_0 /B^2_0$, $\alpha = \frac{1}{\gamma -1}$ and $Y_0(u)$ is the mound height profile expressed in scaled flux coordinate $u$.

We adopt the following set of boundary conditions :
\begin{enumerate}
\item
$\psi = \psi_{\rm d}$ at $z=0$ $\forall \; r$. $\psi_{\rm d}=\frac{1}{2}B_0r^2$ is the initial guess $\psi$ which is the $\psi$  for a uniform field $\mathbf{B}=B_0 \boldsymbol{\hat{z}}$ (approximation of dipolar field in the polar cap region).  This is the line-tying condition of the field at the base \citep{hameury83}. 

\item
$\psi = \psi_{\rm d}$ at $r=0 \; \forall \; z$.

\item
$\psi = \psi_{\rm d} \; \mbox{at} \; r \rightarrow \infty \; \forall \; z$. Field is undistorted for $r >> R_p$. Ideally one should keep the $r=r_{\rm max}$ boundary at infinity to fully capture the  distortion of the field lines resulting from the lateral pressure of the confined plasma. To implement this numerically is impractical. In our GS-solver the  boundary along the radial direction is chosen at $r_{\rm max}=r_p$ where $\psi = \psi _d$ is fixed as the boundary condition. The mound height profiles considered for the GS-solver fall off with radius vanishing at $r=r_p$. Thus at $r=r_p$ boundary there is very little deviation from initial unloaded field configuration. Possible limitations of this are discussed in Sec. ~\ref{solval}.

\item
$\psi = \psi_{\rm d} \; \mbox{at} \; z \rightarrow \infty$. Field is undistorted for $z \rightarrow \infty$. Setting boundary height ($z_{\rm top}$) too close to the maximum height of the mound ($Z_{\rm c}$) affects the solution and gives incorrect result. It is seen that setting $z_{\rm top} \geq 1.5 Z_{\rm c}$ is sufficient for the solution to be stable. Further change in boundary height ($z_{\rm top}$) does not change the solution significantly.

\end{enumerate}

\begin{figure}
	\centering
	\includegraphics[width = 6cm, height = 4.8cm,keepaspectratio] {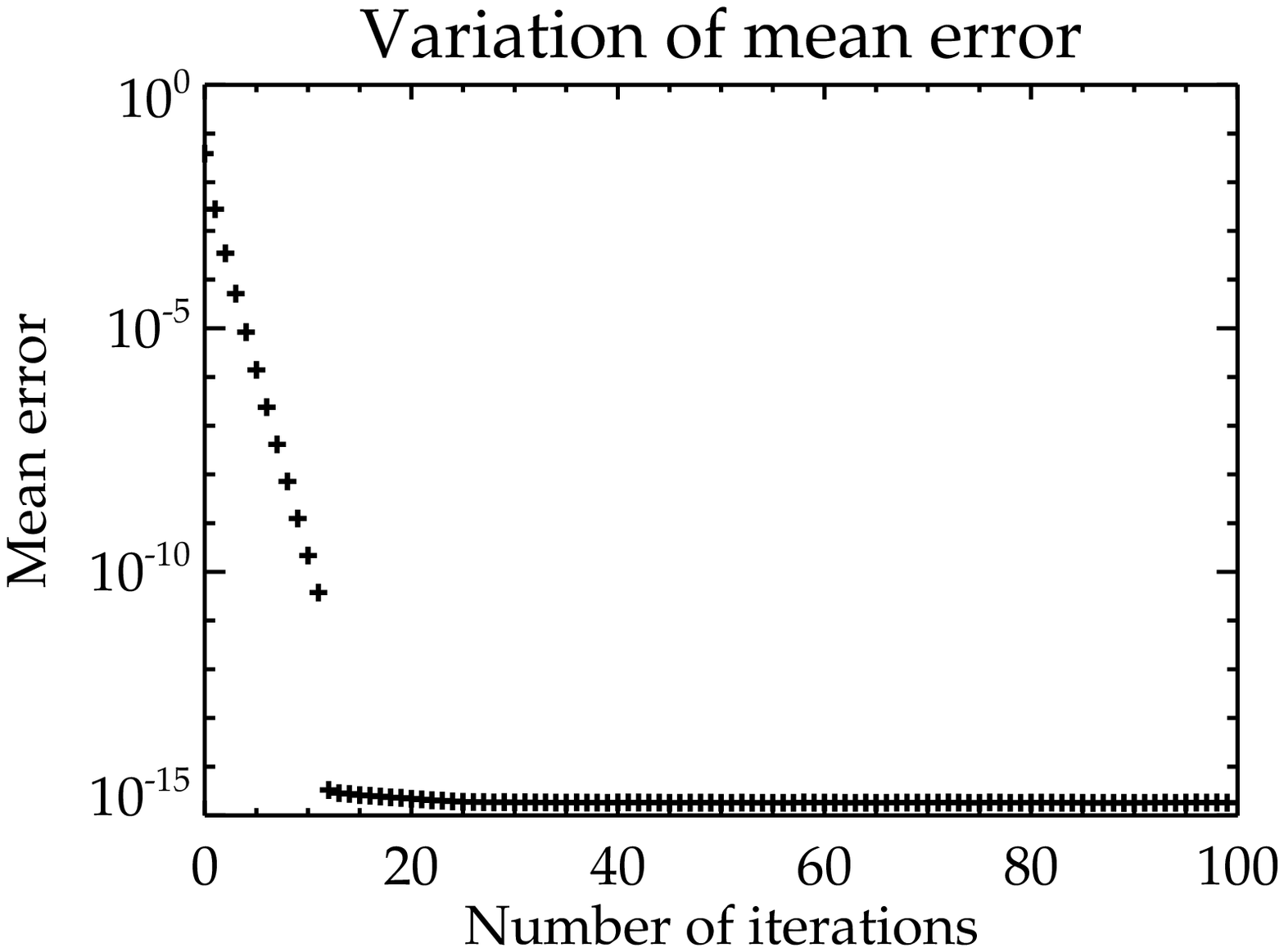}
	\caption{ \small Mean error vs number of iterations of GS-solver till 100 iterations. Convergence (eq. ~\ref{eqconv}) was reached after 9 steps.}
	\label{errfig}
\end{figure}

We adopt a numerical scheme similar to the one followed by \citet{mouschovias74} and \cite{melatos04}. The density is determined from the mound height profile and eq. ~(\ref{rhoeq}).  A two state iterative scheme using an outer under relaxation scheme and an inner Successive Over Relaxation (SOR) method using Chebyshev acceleration \citep{press} is adopted to tackle the non-linear source term on the RHS. We start with a initial guess solution ($\psi _{\rm ini} = \psi _{\rm d}$). SOR loop converges when $\xi \leq \epsilon _{\rm sor}  \xi _{\rm ini}$ where $\xi$ is the residue of differenced equation (LHS-RHS) \citep{press}, $\xi _{\rm ini}$ is the initial residue at the start of the G-S iterations and $\epsilon _{\rm sor}$ is the error tolerance for the SOR scheme. The solution  from the SOR at the $i^{th}$ G-S iteration is then evolved by an under-relaxation scheme
\begin{equation}\label{eqconv}
\left. \begin{aligned} \Delta &= \psi ^i - \psi ^{i-1} \\
\psi^i_{\rm new} &= \psi^i  - \zeta \Delta  
\end{aligned}
\right \}
\end{equation}
where $\zeta \leq 1$ is the under-relaxation parameter. The convergence is considered achieved when maximum error at each grid point is reduced below threshold
\begin{equation}
\frac{\psi^i - \psi ^{i-1}}{\psi^{i-1}} < \epsilon _{\rm GS}
\end{equation}
Smaller values of $\zeta$ gave faster convergence but convergence rate did not improve much with $\zeta$ below an optimum value $ \sim 0.01$. The convergence rate depends on the model being used. Cases for which physical parameters (magnetic field, pressure etc) are near regions where the solution does not exist (see Sec. ~\ref{solval}), the G-S loop takes longer time to converge. Otherwise convergence is reached within less than 100 steps. Fig. \ref{errfig} shows mean error ( $\displaystyle\sum _{\rm N_x, N_y} \frac{\Delta}{\rm N_x \rm N_y}$) at each step of iteration for 100 steps. The convergence was reached after 9 steps in this case. The error tolerance limits were usually set at $\epsilon _{\rm sor} = 10^{-8}$ and $\epsilon _{\rm GS} = 10^{-7}$ for a grid of $1024 \times 1024$. 

\subsection{Testing the GS-solver}
Our GS-solver was tested by comparing results with analytical solutions of the G-S equation. We solved the Soloviev equation 
\begin{equation}
\Delta ^2 \psi = r^2
\end{equation}
whose analytical solution is $\psi _{\rm ana}= \frac{1}{2} r^2 z^2$. The maximum absolute difference of the solution from the GS-solver and the analytic expression was 
\begin{equation}
|\frac{\psi _{\rm num} - \psi _{\rm ana}}{\psi _{\rm ana}}| \leq 9.85 \times 10^{-5}
\end{equation}
which we consider acceptable. 

We also tested the equilibria by using the MHD code PLUTO \citep{andrea07} and checking if the solution is stationary with time. The typical Alfv\'en time scale of the accreted mound with a scale length $L_0 = 1$~km,  density $\rho_0 = 10^6 \mbox{~g cm}^{-3}$ and magnetic field $B_0 = 10^{12}$~G is $t_A = L_0/V_A \sim 3.5 \times 10^{-4}$~s ($V_A$ is the Alfv\'en velocity). Solution from the GS-solver was put as initial condition in PLUTO with a simulation region located inside the density mound. At the boundary, values of the physical parameters (pressure,magnetic field, density) were kept fixed to the initial value obtained from the G-S solution. The simulation was run for several Alfv\'en times and the solution was found to be stationary. This confirms that the solution obtained from our GS-solver represents an equilibrium.

\section{Geometry of hotspot with respect to line of sight (los)} \label{losgeo}
\begin{figure}
	\centering
	\includegraphics[width = 7cm, height = 7cm,keepaspectratio] {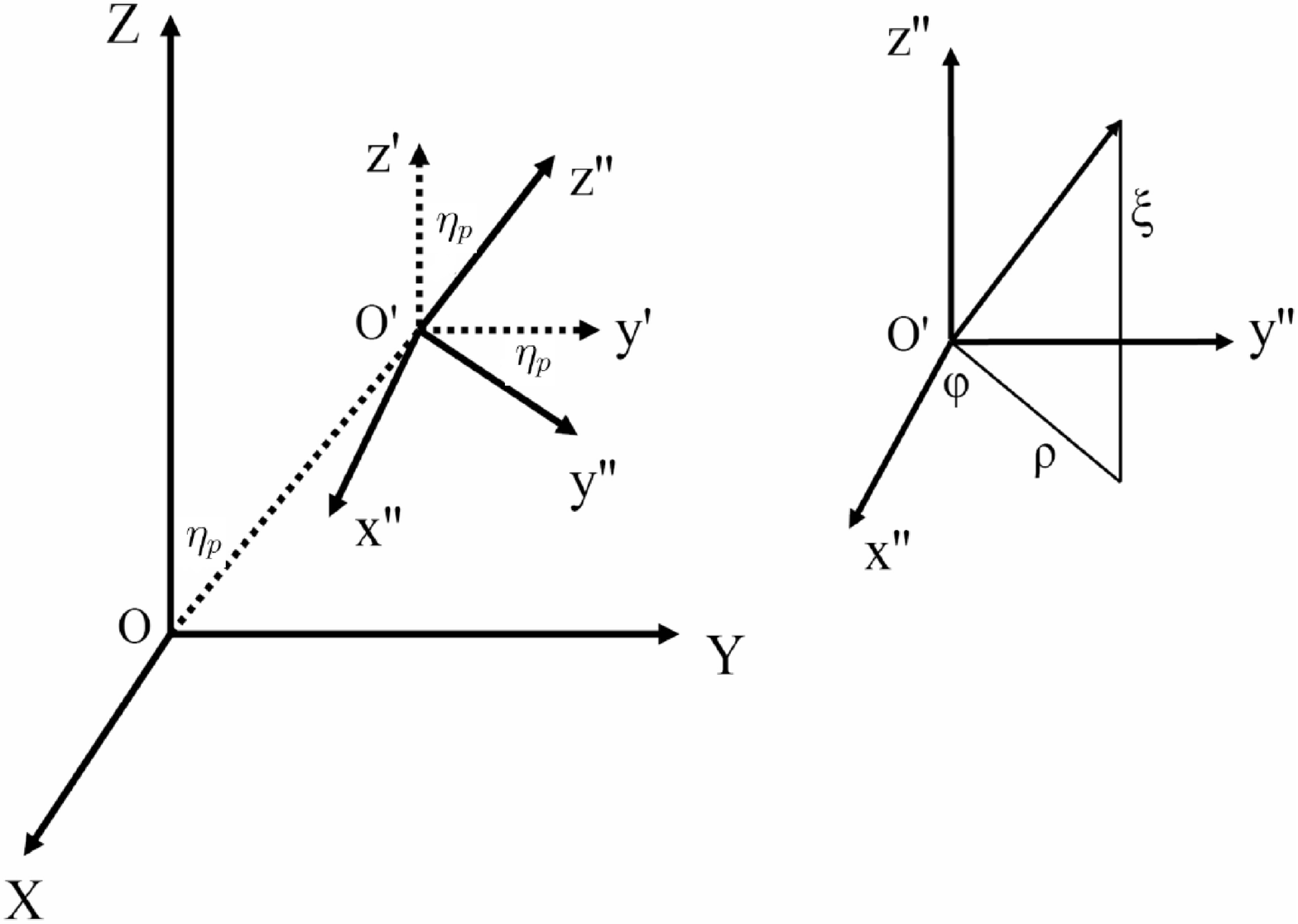}
	\includegraphics[width = 6cm, height = 5.2cm] {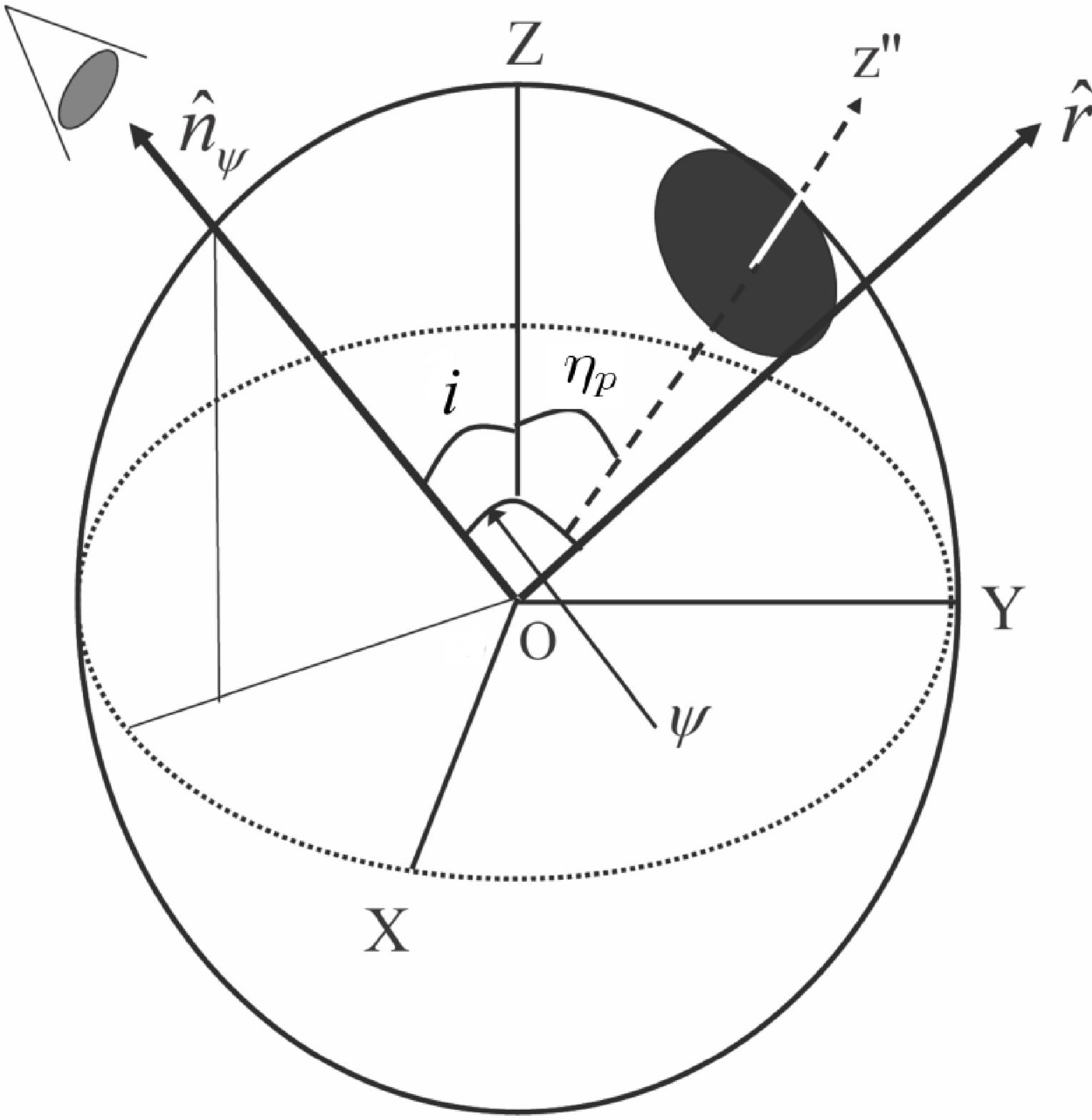}
	\caption{ \small The position and orientation of the coordinate system local to the hotspot (${\rm X}'',{\rm Y}'',{\rm Z}''$) with respect to the global coordinate system (X,Y,Z) defined in the frame of the neutron star. The Z axis is the axis of rotation and ${\rm Z}''$ is along the magnetic pole ${\rm O}'$ (which is also the centre of the hotspot). The magnetic axis (${\rm Z}''$) is at an angle $\eta_p$ to the spin axis (Z). The line of sight makes angle $i$ with the Z axis and $\omega$ (phase) with the  Y axis. In the frame (${\rm X}'',{\rm Y}'',{\rm Z}''$), ($\rho,\phi,\xi$) are the cylindrical coordinates where angle $\phi$ is measured with respect to ${\rm X}''$ axis. X and ${\rm X}''$ axes are in the same direction.}
	\label{hotspot}
\end{figure}

Fig. ~\ref{hotspot} shows the relative position of the hotspot with respect to the observer and the neutron star.  We define the frame ($X,Y,Z$) with origin at the centre of the neutron star (O) such that $\mbox{X}''$ and X are in the same direction and angle between Z,$\mbox{Z}''$ is $\eta_p$. The coordinates of the line of sight (los) with respect to the origin O are ($i$,$\omega$) with $\omega$ measured from the Y-axis. The unit vector along the los is $\hat{\mathbf{n}}_\psi \equiv(\sin i \sin \omega, \sin i \cos \omega, \cos i$). The hotspot lies on the surface  of the star with its centre at $\mbox{O}'$ whose coordinates with respect to the O are ($R_s,\eta_p$) (see Fig. ~\ref{hotspot}), $\mbox{Z}''$ being the normal through $\mbox{O}'$. The G-S computation is done in a cylindrical coordinate system ($\rho,\phi,\xi$) in the frame ($\mbox{X}'',\mbox{Y}'',\mbox{Z}''$) where angle $\phi$ is defined with respect to $\mbox{X}''$. Coordinates of a point in ($\mbox{X}'',\mbox{Y}'',\mbox{Z}''$) frame is ($\rho \cos\phi, \rho\sin\phi,\xi$).

To express the coordinates of a point on the hotpost in the ($X,Y,Z$) frame, we first rotate the ($\mbox{X}'',\mbox{Y}'',\mbox{Z}''$) about $\mbox{X}''$ by angle $\eta_p$ to ($\mbox{X}',\mbox{Y}',\mbox{Z}'$) such that $\mbox{Z}'$ is parallel to Z, and then shift the origin from $\mbox{O}'$ to O by distance $R_s$ along the radial line joining $\mbox{OO}'$. Thus we get the coordinates of a point $\mathbf{r}$ on the hotspot as $\mathbf{r} \equiv \lbrace \rho \cos \phi,\rho \cos \eta_p \sin \phi  + (\xi + R_s) \sin \eta_p,(\xi + R_s) \cos \eta_p  - \rho \sin \eta_p \sin \phi \rbrace$, where $r= \left(\rho ^2 + (\xi + R_s)^2\right)^{1/2}$ is the radial distance of ($x,y,z$) from O.

\begin{figure}
	\centering
	\includegraphics[width = 6cm, height = 6cm,keepaspectratio] {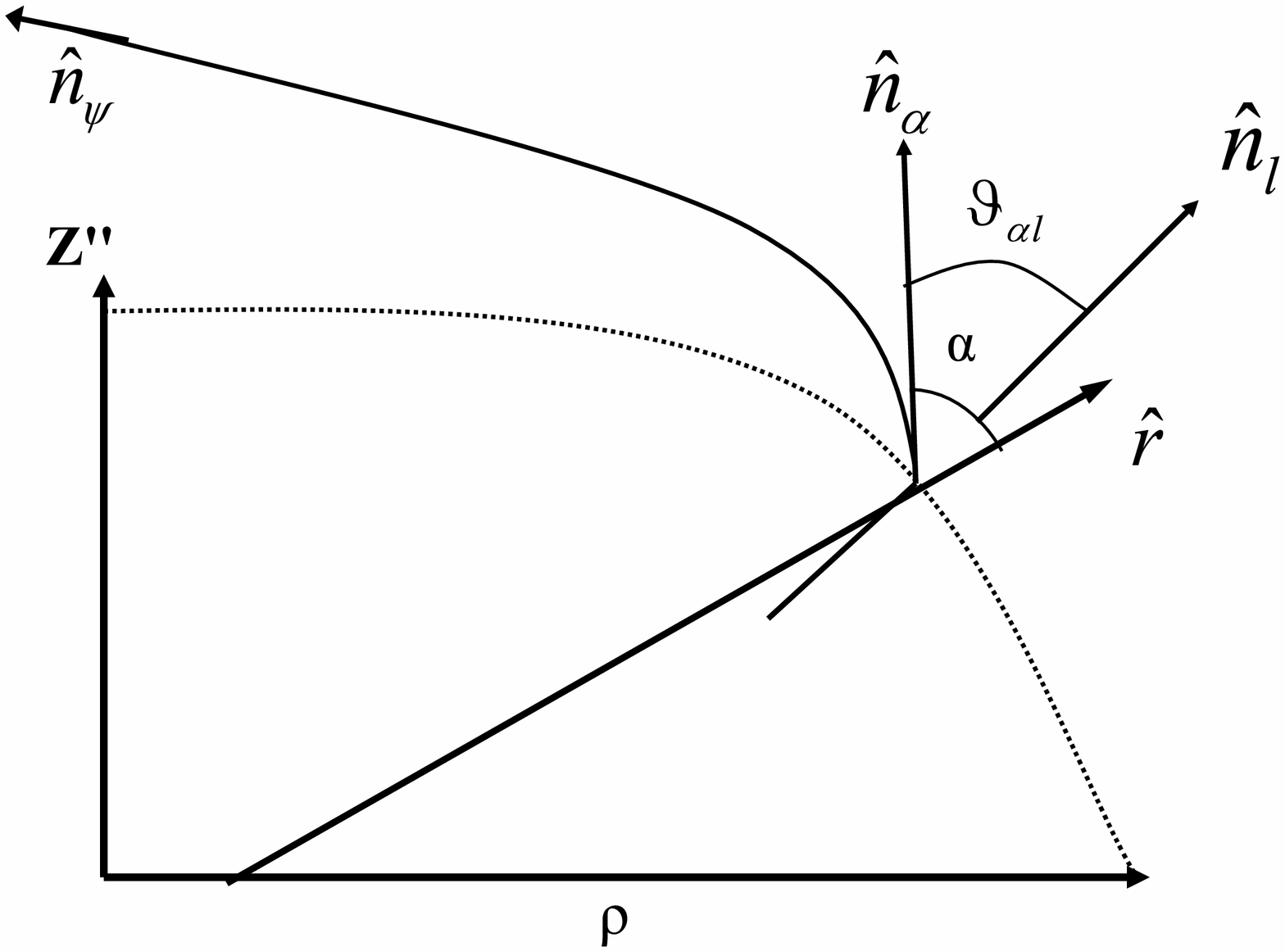}
	\caption{ \small The relative orientation of the local normal $\hat{\mathbf{n}}_l$, local direction of emission $\hat{\mathbf{n}}_\alpha$ of a ray of light and the direction of the bent ray $\hat{\mathbf{n}}_\psi$ which is along the los in the local coordinate system of a representative mound. The line along $\hat{\mathbf{r}}$ is the direction of the radius vector from the star centre ($O$) to the mound (see Fig. ~\ref{hotspot}). A ray of light along $\hat{\mathbf{n}}_\alpha$ makes an angle $\alpha$ with the radius vector $\hat{\mathbf{r}}$ which due to gravitational bending is bent towards $\hat{\mathbf{n}}_\psi$. The local emission direction makes an angle $\theta _{\alpha l}$ with the local normal. The vectors $\hat{\mathbf{r}}, \hat{\mathbf{n}}_\alpha, \hat{\mathbf{n}}_\psi$ and $\hat{\mathbf{n}}_l$ all lie in the same plane.}
	\label{moundlocal}
\end{figure}
The angle between the los ($\hat{\mathbf{n}}_\psi$) and the radius vector to a point on the hotspot is $\cos \psi = \hat{\mathbf{n}}_\psi \cdot \mathbf{r}/r$. The ray coming towards the observer along a los is deviated from its local direction of emission on the neutron star surface ($\hat{\mathbf{n}}_\alpha$, see Fig. ~\ref{moundlocal}) due to gravitational bending of light. The local emission angle ($\alpha$) and the angle between the los and the radius vector $\hat{\mathbf{r}}$ can be related by the approximate formula given by  Beloborodov \citep{beloborodov02,poutanen06} :
\begin{equation}\label{lightbend}
\cos \alpha \simeq u+(1-u)\cos \psi
\end{equation}
where $u=r_s/r$, $r_s$ being the Schwarzschild radius. Since the unit vectors $\hat{\mathbf{n}}_\alpha, \hat{\mathbf{n}}_\psi$ and $\hat{\mathbf{r}}$ lie in the same plane, we can write $\hat{\mathbf{n}}_\alpha \times(\hat{\mathbf{n}}_\alpha \times \hat{\mathbf{r}}) = C  \hat{\mathbf{n}}_\alpha \times(\hat{\mathbf{n}}_\psi \times \hat{\mathbf{r}})$, $C$ being a constant, as the two vectors lie in the same direction differing only in magnitude. Using $\hat{\mathbf{n}}_\alpha \cdot \hat{\mathbf{r}} = \cos \alpha$, the constant $C$ is evaluated to be $C=\sin \alpha/\sin \psi$, for $\alpha \not= 0$. So $\hat{\mathbf{n}}_\alpha$ can be related to $\hat{\mathbf{n}}_\psi$ ({\bf as in \citet{poutanen03}})
\begin{equation}
\hat{\mathbf{n}}_\alpha = \frac{\sin (\psi -\alpha)}{\sin \psi} \hat{\mathbf{r}} + \frac{\sin \alpha}{\sin \psi} \hat{\mathbf{n}}_\psi  \label{nalpha}
\end{equation}
For $\alpha=0$ (which corresponds to $\psi =0$ from eq. ~\ref{lightbend}), we have $\hat{\mathbf{n}}_\alpha=\hat{\mathbf{n}}_\psi=\hat{\mathbf{r}}$. Numerical computation of $\sin (\psi-\alpha)/\sin \psi$ and $\sin \alpha/\sin \psi$ leads to large errors in the limit $(\psi,\alpha) \rightarrow 0$. We expand eq. (~\ref{lightbend}) for small $\alpha$ and $\psi$ : $\alpha=(\sqrt{1-u})\psi$. Using this, we get the following approximate relations
\begin{eqnarray}
\frac{\sin(\psi-\alpha)}{\sin \psi}&\simeq&u+\cos\psi\sqrt{1-u}(\sqrt{1-u}-1) \label{coeff1} \\
\frac{\sin\alpha}{\sin\psi} &\simeq&\sqrt{1-u} \label{coeff2}
\end{eqnarray}
The errors in the approximate form of eq. ~(\ref{nalpha}) for small ($\psi$,$\alpha$) are less than 3\% for eq. ~(\ref{coeff1}) and 1.5\% for eq. ~(\ref{coeff2}) for $\psi \leq 25^\circ$. 

The angle between the local direction of emission ($\hat{\mathbf{n}}_\alpha$) and the local magnetic field ($\hat{\mathbf{b}}$) is required to determine the angular dependence of the width and relative intensity of the cyclotron resonance scattering features (CRSF) (Sec. ~\ref{modelling_cyclotron_spectra}). From eq. ~(\ref{nalpha}) and unit vector along local magnetic field  $\hat{\mathbf{b}} = b_\rho \hat{\rho} + b_\xi \hat{\xi}$ we get the angle $\theta _{\alpha b}$ between $\hat{\mathbf{n}}_\alpha$ and $\hat{\mathbf{b}}$ as :
\begin{equation}
\cos \theta_{\alpha b} = \hat{\mathbf{n}}_\alpha \cdot \hat{\mathbf{b}} = C_1 \cos \theta _{rb} + C_2 \cos \theta _{\psi b}\label{thetab}
\end{equation}
where $\cos \theta _{rb} = \hat{\mathbf{r}}\cdot\hat{\mathbf{b}}= (\rho b_\rho + (\xi + R_s)b_\xi)/r$, and $C_1$ and $C_2$ are the coefficients of $\hat{\mathbf{r}}$ and $\hat{\mathbf{n}}_\psi$ respectively from eq. ~(\ref{nalpha}), eq. ~(\ref{coeff1}) and eq. ~(\ref{coeff2}) whichever is applicable. 
\begin{equation*}
\begin{aligned}
&\cos \theta_{\psi b}=\hat{\mathbf{n}}_\psi \cdot \hat{\mathbf{b}}\\
&= b_\rho [\sin i \sin \omega \cos \phi +(\sin i \cos \omega \cos \eta_p - \cos i \sin \eta_p)\sin \phi] \\
& + b_\xi[\sin i \cos \omega \sin \eta_p + \cos i \cos \eta_p]
\end{aligned}
\end{equation*}

The angle between $\hat{\mathbf{n}}_\alpha$ and the local normal ($\hat{\mathbf{n}}_l$) is required to evaluate the flux from a local area element (eq. ~\ref{fluxfunction}) for a given intensity profile. To find the unit vector along the local normal to the mound ($\hat{\mathbf{n}}_l$) we first fit the top profile of the mound ($\xi_{\rm top}=f(\rho)$ \footnotemark \footnotetext{$\rho$ in this section is the radial coordinate in the hotspot local frame, not to be confused with density.}) with a polynomial function of $\rho$ (the order of the polynomial is chosen to keep errors to less than 5\%). From the slope ($ m_s=d\xi _{\rm top}/d\rho$) we find the normal to the mound in the local frame ($\mbox{X}'',\mbox{Y}'',\mbox{Z}''$)  and transform it to the ($X,Y,Z$) coordinate to get $\hat{\mathbf{n}}_l \equiv \lbrace -\sin \theta _s \cos \phi, -\sin \theta _s \cos \eta_p \sin \phi  + \cos \theta _s \sin \eta_p, \cos \theta _s \cos \eta_p  +  \sin \theta _s \sin \eta_p \sin \phi \rbrace$, where $\sin \theta _s = \frac{m_s}{\sqrt{1+m_s^2}}$ and $\cos \theta _s = \frac{1}{\sqrt{1+m_s^2}}$. Thus we get the angle $\theta _{\alpha l}$ between $\hat{\mathbf{n}}_\alpha$ and $\hat{\mathbf{n}}_l$ as :
\begin{equation}
\cos \theta_{\alpha l} = \hat{\mathbf{n}}_\alpha \cdot \hat{\mathbf{n}}_l = C_1 \cos \theta _{rl} + C_2 \cos \theta _{\psi l}\label{thetal}
\end{equation}
where again $C_1$ and $C_2$ are the coefficients of $\hat{\mathbf{r}}$ and $\hat{\mathbf{n}}_\psi$ respectively from eq. ~(\ref{nalpha}), eq. ~(\ref{coeff1}) and eq. ~(\ref{coeff2}) whichever is applicable. $\cos \theta _{rl} = \hat{\mathbf{r}}\cdot\hat{\mathbf{n}}_l= (\xi + R_s -m_s \rho)/(r\sqrt{1+m_s^2})$ and 
\begin{equation*}
\begin{aligned} &\cos \theta_{\psi l}=\hat{\mathbf{n}}_\psi \cdot \hat{\mathbf{n}}_l \\
&= \frac{m_s}{\sqrt{1+m_s^2}} [(\cos i \sin \eta_p - \sin i \cos \omega \cos \eta_p)\sin \phi \\
& - \sin i \sin \omega \cos \phi] + \frac{1}{\sqrt{1+m_s^2}}[\sin i \cos \omega \sin \eta_p + \cos i \cos \eta_p]
\end{aligned}
\end{equation*}

\def\apj{ApJ}%
\def\mnras{MNRAS}%
\def\aap{A\&A}%
\def\apjl{ApJ}
\def\physrep{PhR}
\def\apjs{ApJS}
\def\pasa{PASA}
\def\pasj{PASJ}

\bibliographystyle{mn}
\bibliography{dipanjanbib}

\end{document}